\documentclass[aps,prb,twocolumn,groupedaddress,longbibliography]{revtex4-1}

\usepackage{amsmath}
\usepackage{amssymb}
\usepackage{graphicx}
\usepackage{subcaption}
\usepackage{xcolor}
\usepackage[colorlinks=true,
pdfborder={0 0 0},
linkcolor=blue,
citecolor=black]{hyperref}

\begin{document}


\title{Spin-Dependent Electron Transmission Model for Chiral Molecules in Mesoscopic Devices}


\author{Xu~Yang}
\email[]{xu.yang@rug.nl}
\affiliation{Zernike Institute for Advanced Materials, University of Groningen, NL-9747AG Groningen, The Netherlands}

\author{Caspar~H.~van~der~Wal}
\affiliation{Zernike Institute for Advanced Materials, University of Groningen, NL-9747AG Groningen, The Netherlands}

\author{Bart~J.~van~Wees}
\affiliation{Zernike Institute for Advanced Materials, University of Groningen, NL-9747AG Groningen, The Netherlands}


\date{\today}

\begin{abstract}

Various device-based experiments have indicated that electron transfer in certain chiral molecules may be spin-dependent, a phenomenon known as the Chiral Induced Spin Selectivity (CISS) effect. However, due to the complexity of these devices and a lack of theoretical understanding, it is not always clear to what extent the chiral character of the molecules actually contributes to the magnetic-field-dependent signals in these experiments. To address this issue, we report here an electron transmission model that evaluates the role of the CISS effect in two-terminal and multi-terminal linear-regime electron transport experiments. Our model reveals that for the CISS effect, the chirality-dependent spin transmission is accompanied by a spin-flip electron reflection process. Furthermore, we show that more than two terminals are required in order to probe the CISS effect in the linear regime. In addition, we propose two types of multi-terminal nonlocal transport measurements that can distinguish the CISS effect from other magnetic-field-dependent signals. Our model provides an effective tool to review and design CISS-related transport experiments, and to enlighten the mechanism of the CISS effect itself.
 
\end{abstract}

\maketitle

\section{Introduction \label{intro}}

Developments in the semiconductor industry have allowed integrated circuits to rapidly shrink in size, reaching the limit of conventional silicon-based electronics. One idea to go beyond this limit is to use the spin degree-of-freedom of electrons to store and process information (spintronics).\cite{wolf2001spintronics} A spintronic device usually contains two important components: a spin injector and a spin detector, through which electrical or optical signals and spin signals can be interconverted. Conventionally, this conversion is done with bulky solid-state materials, but the recently discovered Chiral Induced Spin Selectivity (CISS) effect suggests that certain chiral molecules or their assemblies are capable \textcolor{black}{of} generating spin signals as well. This effect describes that electrons acquire a spin polarization while being transmitted through certain chiral (helical) molecules. Notably, experimental observations of the CISS effect suggest its existence, but complete theoretical insight in its origin is still lacking.\cite{naaman2012chiral,naaman2015spintronics} The CISS effect is thus not only relevant for spintronic applications, but also fundamentally interesting.

The CISS effect has been experimentally reported in chiral (helical) systems ranging from large biological units such as dsDNA \cite{gohler2011spin,xie2011spin} to small molecules such as helicenes. \cite{kiran2016helicenes,kettner2018chirality} Typically, these experiments can be categorized into either electron photoemission experiments \cite{ray1999asymmetric,carmeli2002magnetization,ray2006chirality,gohler2011spin,mishra2013spin,nino2014enantiospecific,kettner2015spin,kettner2018chirality} or magnetotransport measurements. \cite{xie2011spin,michaeli2017new,dor2013chiral,mondal2015chiral,varade2018bacteriorhodopsin,bloom2016spin,mathew2014non,alam2015spin,kiran2016helicenes} The latter, in particular, are usually based on solid-state devices and are of great importance to the goal of realizing chiral-molecule-based spintronics. Important to realize, in such devices, the CISS-related signals may often be overshadowed by other spurious signals that arise from magnetic components of the devices. Therefore, it is essential to understand the exact role of chiral molecules in these devices and to distinguish between the CISS-related signals and other magnetic-field-dependent signals. However, this has not been addressed, and an effective tool to perform such analyses is still missing.

We provide here a model that is based on the Landauer-B\"uttiker-type of analysis of linear-regime electron transmission and reflection. Unlike other theoretical works,\cite{gutierrez2012spin,yeganeh2009chiral,gersten2013induced,medina2015continuum,michaeli2015origin,maslyuk2018enhanced,guo2012spin,guo2014spin,pan2016spin} our model is derived from symmetry theorems that hold for electrical conduction in general and does not require any assumptions about the CISS effect on a molecular level. With this model, we quantitatively demonstrate how the CISS effect leads to spin injection and detection in linear-regime devices, and analyze whether typical two-terminal and four-terminal measurements are capable of detecting the CISS effect in the linear regime. 

\section{Model \label{model}}

We consider a solid-state device as a linear-regime circuit segment whose constituents are described by the following set of rules: 
\begin{itemize}
	\item A \textit{contact} (pictured as a wavy line segment perpendicular to the current flow, see e.g. Fig.~\ref{fig:2Tcfm}) is described as an electron reservoir with a well-defined chemical potential $\mu$, which determines the energy of the electrons that leave the reservoir. A reservoir absorbs all incoming electrons regardless of its energy or spin;	
	
	\item A \textit{node} (pictured as a circle, see later figures for four-terminal geometries) is a circuit constituent where chemical potentials for charge and spin are defined. It is described by two chemical potentials $\mu_\rightarrow$ and $\mu_\leftarrow$, one for each spin species with the arrows indicating the spin orientations. At a node a spin accumulation $\mu_s$ is defined ($\mu_s=\mu_\rightarrow - \mu_\leftarrow$). Inside a node the momentum of electrons is randomized, while the spin is preserved. \textcolor{black}{The function and importance of the node will be further addressed in the discussion section};

	\item A \textit{CISS molecule} (pictured as a helix, color-coded and labeled for its chirality, see e.g. Fig.~\ref{fig:2Tcfm}), a \textit{ferromagnet} (a filled square, see e.g. Fig.~\ref{fig:2Tcfm}), \textcolor{black}{and} a \textit{non-magnetic barrier} (a shaded rectangle, see e.g. Fig.~\ref{fig:matrices}) are viewed as \textcolor{black}{two-terminal} circuit constituents with energy-conserving electron transmissions and reflections. \textcolor{black}{Each of them} is described by a set of (possibly spin-dependent) transmission and reflection probabilities;
	
	\item The above constituents are connected to each other via \textit{transport channels} (pictured as line segments along the current flow, see e.g. Fig.~\ref{fig:2Tcfm}), \textcolor{black}{in which} both the momentum and the spin of electrons are preserved.
\end{itemize}

\textcolor{black}{Before proceeding with introducing the model, we would like to highlight the important role of dephasing in the generation of the CISS effect. In a fully phase-coherent two-terminal electron transport system, time-reversal symmetry prohibits the production of spin polarization by a charge current.\cite{bardarson2008proof} Consequently, a CISS molecule requires the presence of dephasing in order to exhibit a CISS-type spin-polarizing behavior. The necessity of dephasing has already been addressed by other theoretical works.\cite{matityahu2016spin,matityahu2017spin,guo2014contact} Here we emphasize that the Landauer-Buttiker type of analysis, on which our model is based, does not require phase coherence.\cite{buttiker1986role,buttiker1985generalized} Moreover, dephasing can be naturally provided by inelastic processes such as electron-phonon interactions under experimental conditions. Therefore, it is reasonable to assume that a CISS molecule is able to generate a spin polarization in a linear-regime circuit segment, and our discussions focus on whether this spin polarization can be detected as a charge signal.}

\textcolor{black}{In the following part of this article}, we first derive a key transport property of CISS molecules and then introduce a matrix formalism to quantitatively describe linear-regime transport devices. \textcolor{black}{Later, in the discussion section, we provide analyses for a few experimental circuit geometries.}

\subsection{Reciprocity and spin-flip reflection by chiral molecules}

In order to characterize the CISS effect without having to understand it on a molecular level, we look at universal rules that apply to any conductor in the linear regime, namely the law of charge conservation and the reciprocity theorem.

The reciprocity theorem states that for a multi-terminal circuit segment in the linear regime, the measured conductance remains invariant when an exchange of voltage and current contacts is accompanied by a reversal of magnetic field $H$ and magnetization $M$ (of all magnetic components).\cite{buttiker1986four,buttiker1988symmetry} Mathematically we write
\begin{equation}\label{eqn:recip}
G_{ij,mn}(H, M)=G_{mn,ij}(-H, -M),
\end{equation}
where $G_{ij,mn}$ is the four-terminal conductance measured using current contacts $i$ and $j$ and voltage contacts $m$ and $n$. In two-terminal measurements, this theorem reduces to
\begin{equation}\label{eqn:recip2T}
G_{ij}(H, M) = G_{ij}(-H, -M),
\end{equation}
meaning that the two-terminal conductance remains constant under magnetic field and magnetization reversal. This theorem emphasizes the universal symmetry independent of the microscopic nature of the transport between electrical contacts. It is valid for any linear-regime circuit segment regardless of the number of contacts, or the presence of inelastic scattering events.\cite{buttiker1988symmetry}

\begin{figure}[h]
	\includegraphics[width=\linewidth]{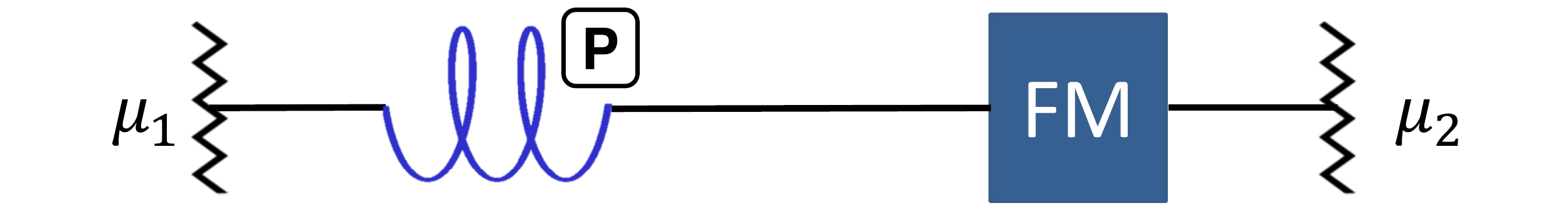}
	\caption{A two-terminal circuit segment with a \textit{P}-type CISS molecule and a ferromagnet between contacts $1$ and $2$ (with chemical potentials $\mu_1$ and $\mu_2$). The notion \textit{P}-type represents the chirality of the molecule and indicates that it allows higher transmission for spins \textit{parallel} to the electron momentum. (The opposite chirality allows higher transmission for spins \textit{anti-parallel} to the electron momentum, and is denoted as \textit{AP}-type.) The ferromagnet (\textit{FM}) is assumed to allow higher transmission of spins parallel to its magnetization direction, which can be controlled to be either parallel or anti-parallel to the electron transport direction.}\label{fig:2Tcfm}
\end{figure}

By applying the reciprocity theorem to a circuit segment containing CISS molecules, one can derive a special transport property of these molecules. For example, in the two-terminal circuit segment shown in Fig.~\ref{fig:2Tcfm}, the reciprocity theorem requires that the two-terminal conductance remains unchanged when the magnetization direction of the ferromagnet is reversed. Since the two-terminal conductance is proportional to the transmission probability between the two contacts (Landauer-B\"uttiker),\cite{buttiker1985generalized} this requirement translates to 
\begin{equation}\label{eqn:recp1}
T_{\textcolor{black}{21}}(\Rightarrow)=T_{\textcolor{black}{21}}(\Leftarrow),
\end{equation}
where \textcolor{black}{$T_{21}$ describes the transmission probability of electrons injected from contact $1$ to reach contact $2$, and} $\Rightarrow$ and $\Leftarrow$ indicate the magnetization directions of the ferromagnet. This requirement gives rise to a necessary spin-flip process associated with the CISS molecule, as described below.

For ease of illustration, we assume an ideal case where both the ferromagnet and the CISS molecule allow a 100\% transmission of the favored spin and a 100\% reflection of the other (the general validity of the conclusions is addressed in Appendix~\ref{app:sprefl}). We consider electron transport from contact $1$ to contact $2$ (see Fig.~\ref{fig:2Tcfm}) and compare the two transmission probabilities $T_{\textcolor{black}{21}}(\Rightarrow)$ and $T_{\textcolor{black}{21}}(\Leftarrow)$. For $T_{\textcolor{black}{21}}(\Rightarrow)$, the \textit{P}-type CISS molecule (favors spin parallel to electron momentum, see figure caption) allows the transmission of spin-right electrons, while it reflects spin-left electrons back to contact $1$. At the same time, the ferromagnet is magnetized to also only allow the transmission of spin-right electrons. Therefore, all spin-right (and none of the spin-left) electrons can be transmitted to contact $2$, giving $T_{\textcolor{black}{21}}(\Rightarrow)=0.5$. As for $T_{\textcolor{black}{21}}(\Leftarrow)$, while the \textit{P}-type CISS molecule still allows the transmission of spin-right electrons, the ferromagnet no longer does. It reflects the spin-right electrons towards the CISS molecule with their momentum anti-parallel to their spin. As a result, these electrons are reflected by the CISS molecule and are confined between the CISS molecule and the ferromagnet. This situation gives $T_{\textcolor{black}{21}}(\Leftarrow)=0$, which is not consistent with Eqn.~\ref{eqn:recp1}. In order to satisfy Eqn.~\ref{eqn:recp1}, i.e. to have $T_{\textcolor{black}{21}}(\Leftarrow)=0.5$, a spin-flip process has to take place for the spin-right electrons, so that they can be transmitted to contact $2$ through the ferromagnet. Such a process does not exist for the ideal and exactly aligned ferromagnet. Therefore, a spin-flip electron reflection process must exist for the CISS molecule. Further analysis (see Appendix~\ref{app:sprefl}) shows that such a spin-flip reflection process completely meets the broader restrictions from Eqn.~\ref{eqn:recip2T}. In addition, the conclusion that a spin-flip reflection process must exist is valid for general cases where the ferromagnet and the CISS molecule are not ideal (see Appendix~\ref{app:sprefl}).

In these derivations, the only assumption regarding the CISS molecule is that it allows higher transmission of one spin than the other, which is a conceptual description of the CISS effect itself. Therefore, the spin-flip reflection process has to be regarded as an inherent property of the CISS effect in a linear-regime transport system, and this is guaranteed by the universal symmetry theorems of electrical conduction.\cite{buttiker1986four,buttiker1988symmetry}

\subsection{Matrix formalism and barrier-CISS center-barrier (BCB) model for CISS molecules}

\begin{figure}[h]
	\includegraphics[width=\linewidth]{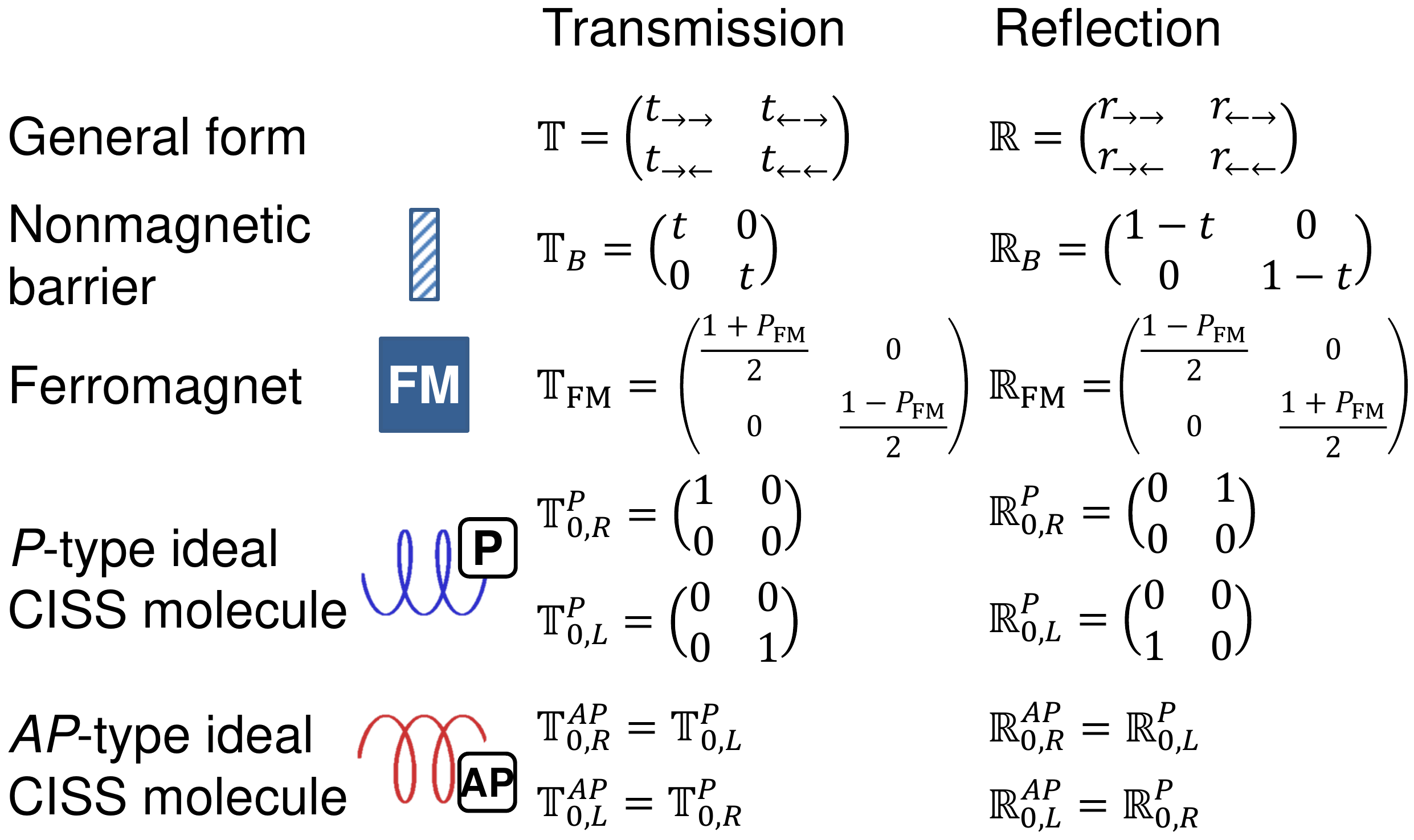}
	\caption{Transmission and reflection matrices ($\mathbb{T}$ and $\mathbb{R}$) for a non-magnetic barrier (subscript \textit{B}, here we use the term \textit{barrier}, but it refers to any circuit constituent with spin-independent electron transmission and reflection), a ferromagnet (subscript \textit{FM}), and ideal \textit{P}-type (superscript \textit{P}) and \textit{AP}-type (superscript \textit{AP}) CISS molecules. For the CISS molecules the subscripts \textit{R} (right) and \textit{L} (left) denote the direction of the \textcolor{black}{incoming} electron flow, and the indicator $0$ in the subscripts means these matrices are for an ideal cases where all the matrix elements are either $1$ or $0$. \textcolor{black}{The matrices for \textit{AP}-type molecules are derived from those for \textit{P}-type molecules under the assumption that opposite chiral enantiomers are exact mirror images of each other, and therefore selects opposite spins with equal probability.} Each matrix element represents the probability of a spin-dependent transmission or reflection, with the \textcolor{black}{column/row} position indicating the corresponding spin orientations before/after the transmission or reflection (see general form in the top row).}\label{fig:matrices}
\end{figure}

We use matrices to quantitatively describe the spin-dependent transmission and reflection probabilities of CISS molecules and other circuit constituents, as shown in Fig.~\ref{fig:matrices}. At the top of the figure, the general form of these matrices is introduced. Matrix element $t_{\alpha\beta}$ (or $r_{\alpha\beta}$), where $\alpha$ and $\beta$ is either \textit{left} ($\leftarrow$) or \textit{right} ($\rightarrow$), represents the probability of a spin-$\alpha$ electron being transmitted (or reflected) as a spin-$\beta$ electron, and $\alpha \neq \beta$ indicates a spin-flip process. Here $0\leq t_{\alpha\beta}, r_{\alpha\beta} \leq 1$, and the spin orientations are chosen to be either parallel or anti-parallel to the electron momentum in later discussions. Next, the transmission and reflection matrices of a non-magnetic barrier are given. These matrices are spin-independent and are fully determined by a transmission probability $t$ ($0\leq t\leq 1$), which depends on the material and dimensions of the barrier. Here we use the term \textit{barrier}, but it refers to any circuit constituent with spin-independent electron transmission and reflection. In the third row, we show the transmission and reflection matrices of a ferromagnet. These matrices are spin-dependent, and are determined by the polarization $P_{FM}$ ($0 < |P_{FM}|\leq1$) of the ferromagnet. Finally, for \textit{P}-type and \textit{AP}-type CISS molecules, we show here an ideal case where all the matrix elements are either $1$ or $0$. The non-zero off-diagonal terms in the reflection matrices represent the characteristic spin-flip reflections. These ideal CISS molecules are later referred to as \textit{CISS centers}, and will be generalized for more realistic situations.

In accordance with the matrix formalism, we use \textcolor{black}{column vector $\boldsymbol{\mu}=\begin{pmatrix} \mu_{\rightarrow}\\\mu_{\leftarrow} \end{pmatrix}$} to describe chemical potentials, and \textcolor{black}{column vector $\boldsymbol{I}=\begin{pmatrix} I_{\rightarrow}\\I_{\leftarrow}	\end{pmatrix}$} to describe currents, where each vector element describes the contribution from one spin component (indicated by arrow).

\begin{figure}[h]
	\includegraphics[width=\linewidth]{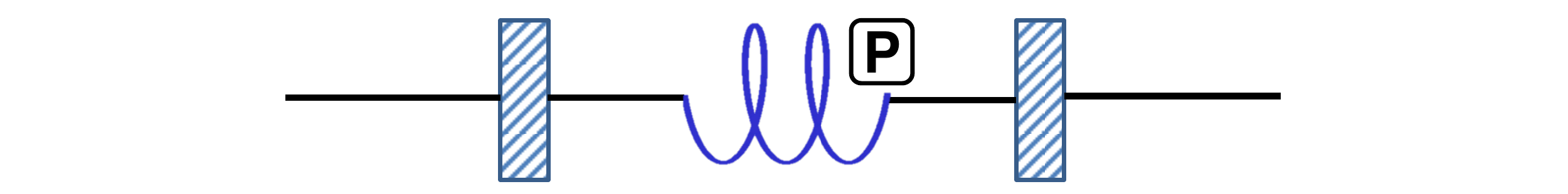}
	\caption{A generalized \textit{Barrier-CISS Center-Barrier (BCB)} model for \textit{P}-type CISS molecules. The ideal, 100\%-spin-selective CISS Center in the middle introduces the directional spin transmission in a CISS molecule, while the two identical non-magnetic barriers (with transmission probability $t$) contribute the non-ideal electron transmission and reflection behavior. The overall transmission and reflection matrices of the entire BCB module are fully determined by $t$ and have all elements taking finite values between $0$ and $1$.}\label{fig:bcb}
\end{figure}

A non-ideal CISS molecule with $C_2$ symmetry (two-fold rotational symmetry with \textcolor{black}{an} axis perpendicular to the electron transport path) can be modeled \textcolor{black}{as} a linear arrangement of two identical barriers sandwiching an ideal CISS center, as shown in Fig.~\ref{fig:bcb} (only the \textit{P}-type is shown). In this \textit{Barrier-CISS Center-Barrier (BCB) model} we consider that all spin-dependent linear-regime transport properties of \textcolor{black}{a CISS molecule} exclusively originate from an ideal CISS center inside the molecule, and the overall spin-dependency is limited by the multiple spin-independent transmissions and reflections at other parts (non-magnetic barriers) of the molecule. Therefore, the barrier transmission probability $t$ ($0< t\leq 1$) fully determines the transmission and reflection matrices of the entire BCB molecule, and consequently determines the spin-related properties of the molecule. The use of an identical barrier on each side of the CISS center is to address the $C_2$ symmetry. However, we stress that not all CISS molecules have this symmetry, and the BCB model is still a simplified picture. The model can be further generalized by removing the restriction of the CISS center being ideal, and this case is discussed in Appendix~\ref{app:general}. Despite being a simplified picture, the BCB model captures all qualitative behaviors of a non-ideal CISS molecule, and at the same time keeps quantitative analyses simple. Therefore, we further only discuss the case of \textit{BCB molecules}, instead of the more generalized \textit{CISS molecules}.

\section{Discussions \label{discussion}}

In this section, we use different approaches to separately analyze two-terminal and multi-terminal circuit geometries. \textcolor{black}{For} two-terminal geometries, we evaluate the conductance of the circuit segment by calculating the electron transmission probability $T_{\textcolor{black}{21}}$ between the two contacts. In contrast, for multi-terminal geometries, we take a circuit-theory approach to evaluate the spin accumulation $\mu_s$ at the nodes. 

A major difference between the two approaches is the inclusion of nodes in multi-terminal geometries. \textcolor{black}{In our description, a node is the only location where spin accumulation can be defined. It can be experimentally realized with a diffusive electron transport channel segment that is much shorter (along the electron transport direction) than the spin-diffusion length $\lambda_s$ of the channel material. Due to its diffusive nature, a} node emits electrons to all directions, so it can be considered as a source of electron back-scattering. Notably, adding a node to a near-ideal electron transport channel (with transmission probability close to $1$) significantly alters its electron transmission probability. \textcolor{black}{Nonetheless, this does not affect the validity of our approach} because we only address non-ideal circuit segments where electron back-scattering (reflection) already exists due to other circuit constituents (CISS molecules, ferromagnets, or non-magnetic barriers). Note that even when we discuss the use of ideal CISS molecules or ideal ferromagnets, the entire circuit segment is non-ideal due to the reflection of the rejected spins. 

In the following discussion, we consider only the \textit{P}-type BCB molecule, and we use expressions $\mathbb{T}^P_{R,L}$ and $\mathbb{R}^P_{R,L}$ to describe transmission and reflection matrices of the entire BCB module, where the subscripts consider electron flow directions. The derivations of these matrices can be found in Appendix~\ref{app:BCB}.

\subsection{Two-terminal geometries} \label{sec:2T}

We discuss here two geometries that are relevant for two-terminal magnetoresistance measurements.\cite{prinz1998magnetoelectronics} 

\begin{figure}[h]
	\includegraphics[width=\linewidth]{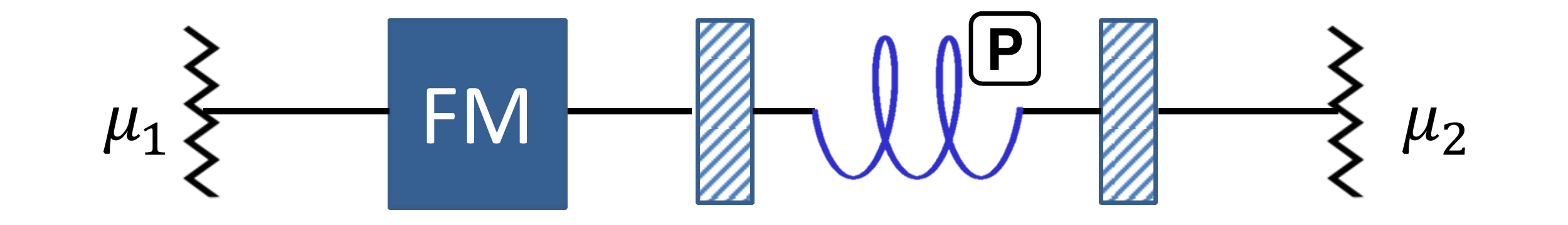}
	\caption{An \textit{FM-BCB} geometry where a ferromagnet and a BCB molecule are connected in series in a two-terminal circuit segment. The magnetization reversal of the ferromagnet does not change the two-terminal conductance.}\label{fig:2Tafm}
\end{figure}

The first is an \textit{FM-BCB} geometry, as shown in Fig.~\ref{fig:2Tafm}. It simulates a common type of experiment where a layer of chiral molecules is sandwiched between a ferromagnetic layer and a normal metal contact. The other side of the ferromagnetic layer is also connected to a normal metal contact (experimentally this may be a wire that connects the sample with the measurement instrument). Due to the spin-dependent transmission of the chiral molecules and the ferromagnet, one might expect a change of the two-terminal conductance once the magnetization of the ferromagnet is reversed. However, this change is not allowed by the reciprocity theorem (Eqn.~\ref{eqn:recip2T}), which can be confirmed with our model, as explained below.

In order to illustrate this, we calculate the electron transmission probabilities between the two contacts for opposite ferromagnet magnetization directions, $T^{FM-BCB}_{\textcolor{black}{21}}(\Rightarrow)$ and $T^{FM-BCB}_{\textcolor{black}{21}}(\Leftarrow)$, where the arrows indicate the magnetization directions.

For the magnetization direction to the right ($\Rightarrow$), we first derive the transmission and reflection matrices with the combined contribution from the ferromagnet and the BCB molecule 
\begin{subequations}\label{eqn:2tmat}
	\color{black}
	\begin{align}
	\begin{split}
	&\mathbb{T}^{FM-BCB}_{21}(\Rightarrow) \\
	=~&\mathbb{T}^P_R \cdot \Bigg( \mathbb{I}+ \mathbb{R}_{FM}(\Rightarrow) \cdot \mathbb{R}^P_R\\ 
	&+ \Big( \mathbb{R}_{FM}(\Rightarrow) \cdot \mathbb{R}^P_R \Big)^2 + \Big( \mathbb{R}_{FM}(\Rightarrow) \cdot \mathbb{R}^P_R \Big)^3 \\
	&+\cdots \Bigg) \cdot \mathbb{T}_{FM}(\Rightarrow) \\
	=~&\mathbb{T}^P_R \cdot \Big( \mathbb{I}-\mathbb{R}_{FM}(\Rightarrow) \cdot \mathbb{R}^P_R \Big)^{-1} \cdot \mathbb{T}_{FM}(\Rightarrow),
	\end{split}
	\end{align}
	\begin{align}
	\begin{split}
	&\mathbb{R}^{FM-BCB}_{11}(\Rightarrow) \\
	=~&\mathbb{R}_{FM}(\Rightarrow)+ \mathbb{T}_{FM}(\Rightarrow) \cdot \Big(\mathbb{I}-\mathbb{R}^P_R \cdot \mathbb{R}_{FM}(\Rightarrow)  \Big)^{-1} \\
	&\cdot \mathbb{R}^P_{R} \cdot \mathbb{T}_{FM}(\Rightarrow),
	\end{split}
	\end{align}
\end{subequations}
where the $\mathbb{I}=\begin{pmatrix}
1&0\\0&1
\end{pmatrix}$ is the identity matrix. The addition of the multiple reflection terms is due to the multiple reflections between the ferromagnet and the BCB molecule. Next, we include the contribution from the contacts and derive the transmission and reflection probabilities \textcolor{black}{accounting for both spins}
\begin{subequations}\label{eqn:2ttrans}
	\color{black}
	\begin{equation}
	T^{FM-BCB}_{21}(\Rightarrow)= \Big(1,1\Big) \mathbb{T}^{FM-BCB}_{21}(\Rightarrow)\begin{pmatrix}
	1/2\\1/2 \end{pmatrix},
	\end{equation}
	\begin{equation}
	R^{FM-BCB}_{11}(\Rightarrow)= \Big(1,1\Big)\mathbb{R}^{FM-BCB}_{11}(\Rightarrow)\begin{pmatrix}
	1/2\\1/2 \end{pmatrix},
	\end{equation}
\end{subequations} 
where the \textcolor{black}{column vector $\begin{pmatrix}
1/2\\1/2 \end{pmatrix}$} describes the \textcolor{black}{normalized} input current from contact $1$ with equal spin-right and spin-left contributions, and the \textcolor{black}{row vector $\Big(1,1\Big)$ is an operator that} describes the absorption of both spins into contact $2$ \textcolor{black}{(calculates the sum of the two spin components)}.

For the opposite magnetization direction ($\Leftarrow$), we change the magnetization-dependent terms in Eqn.~\ref{eqn:2tmat} and Eqn.~\ref{eqn:2ttrans} accordingly. Detailed calculations (see Appendix~\ref{app:BCB}) prove
\begin{equation}
T^{FM-BCB}_{\textcolor{black}{21}}(\Rightarrow) \equiv T^{FM-BCB}_{\textcolor{black}{21}}(\Leftarrow)
\end{equation}
for all BCB transmission probabilities $t$ and all ferromagnet polarizations $P_{FM}$. Therefore, it is not possible to detect any variation of two-terminal conductance in this geometry by switching the magnetization direction of the ferromagnet. In Appendix~\ref{app:BCB} we show that it is also not possible to detect any variation of two-terminal conductance by reversing the current, and that the above conclusions also hold for the more generalized CISS model (Appendix~\ref{app:general}). These conclusions \textcolor{black}{also} agree with earlier reports on general voltage-based detections of current-induced spin signals.\cite{adagideli2006detection} 

\begin{figure}[h]
	\includegraphics[width=\linewidth]{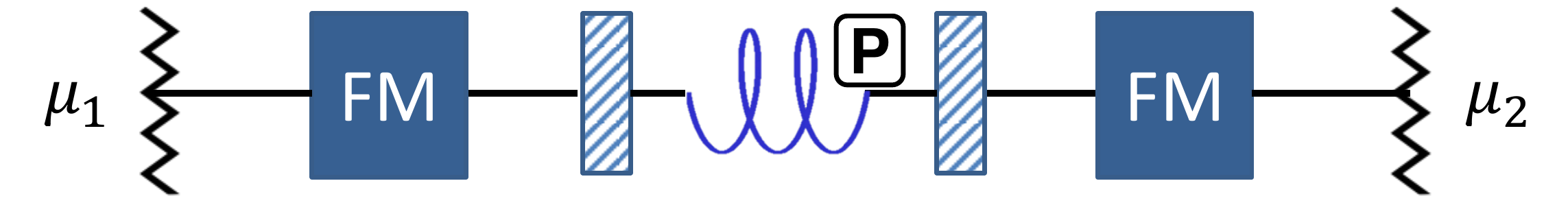}
	\caption{A \textit{Spin Valve} geometry with a BCB molecule placed in between two ferromagnets. Unlike a conventional spin valve, here the magnetization reversal of one ferromagnet does not change the two-terminal conductance due to the presence of the BCB molecule.}\label{fig:2Tsv}
\end{figure}

The second geometry, as shown in Fig.~\ref{fig:2Tsv}, contains two ferromagnets, and is similar to a spin valve. \textcolor{black}{In a conventional spin valve (a non-magnetic barrier sandwiched between two ferromagnets), the magnetization reversal of one ferromagnet leads to a change of the two-terminal conductance\cite{prinz1998magnetoelectronics} (this does not violate the reciprocity theorem since switching one ferromagnet does not reverse all magnetizations of the entire circuit segment), whereas in the geometry shown in Fig.~\ref{fig:2Tsv}, this change does not happen due to the presence of spin-flip electron reflections in the BCB molecule. (See Appendix~\ref{app:BCB} for more details.)} As a result, this geometry is not able to quantitatively measure the CISS effect. We emphasize that here the absence of the spin-valve behavior is unique for the BCB model, which contains an ideal CISS Center. In Appendix~\ref{app:general} we show that a further-generalized CISS model regains the spin-valve behavior. Nevertheless, one cannot experimentally distinguish whether the regained spin-valve behavior originates from the CISS molecule or a normal non-magnetic barrier, and therefore cannot draw any conclusion about the CISS effect. In general, it is not possible to measure the CISS effect in the linear regime using two-terminal experiments. 

\subsection{Four-terminal geometries and experimental designs}

\begin{figure}[h!]
	\includegraphics[width=\linewidth]{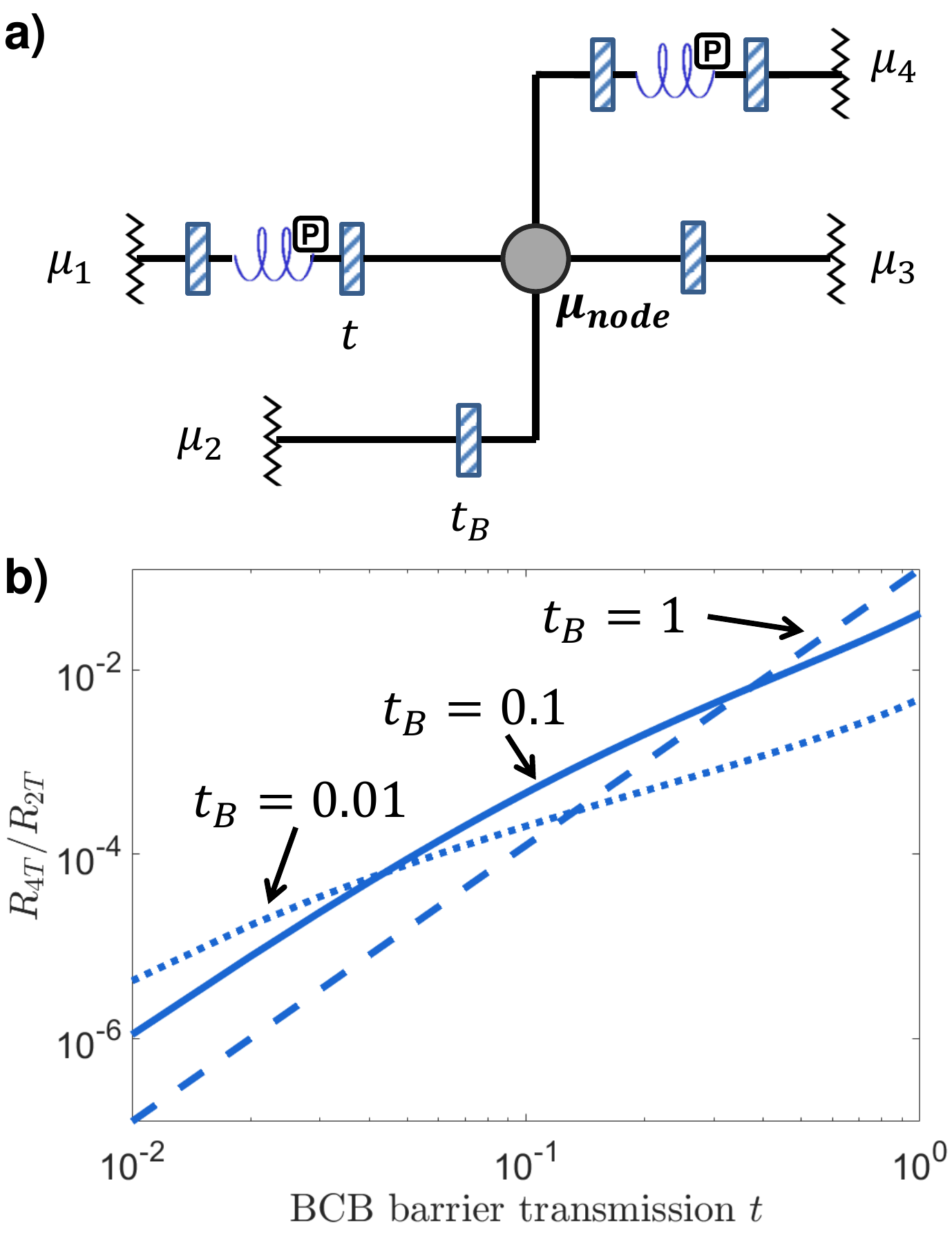}
	\caption{\textbf{a).} A four-terminal geometry that includes a node. Two of the contacts contain BCB molecules, and the other two are coupled to the node via tunnel barriers (with transmission probability $t_B$). The node is characterized by a spin-dependent chemical potential vector \textcolor{black}{$\boldsymbol{\mu_{node}}=\begin{pmatrix} \mu_{node \rightarrow}\\\mu_{node \leftarrow}\end{pmatrix}$}, and each of the four contacts is characterized by a spin-independent chemical potential $\mu_i$, with $i=1, 2, 3, 4$. \textbf{b).} Calculated ratio between four-terminal and two-terminal resistances for this geometry, plotted as a function of $t$ (transmission probability of the barriers in BCB molecules) for three $t_B$ (transmission probability of the barriers at the contacts) values.}\label{fig:4Tcross}
\end{figure}

\begin{figure}[h]
	\includegraphics[width=\linewidth]{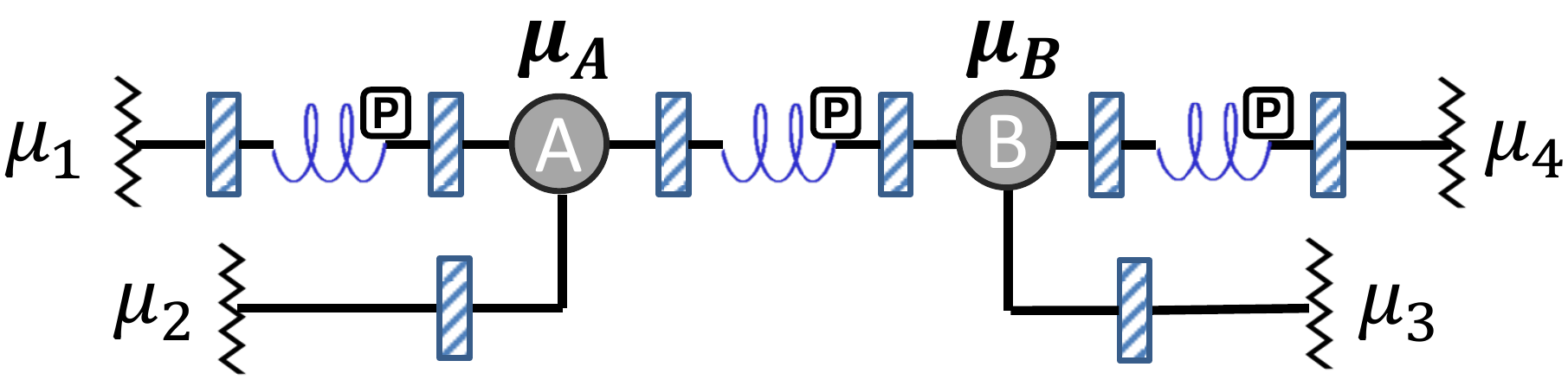}
	\caption{A four-terminal geometry involving two nodes $A$ and $B$, which are connected to each other via a BCB molecule. A spin accumulation difference between the two nodes results in a (charge) chemical potential difference between them, and \textit{vice versa}.}\label{fig:4Tlinear}
\end{figure}

\begin{figure}[h]
	\includegraphics[width=\linewidth]{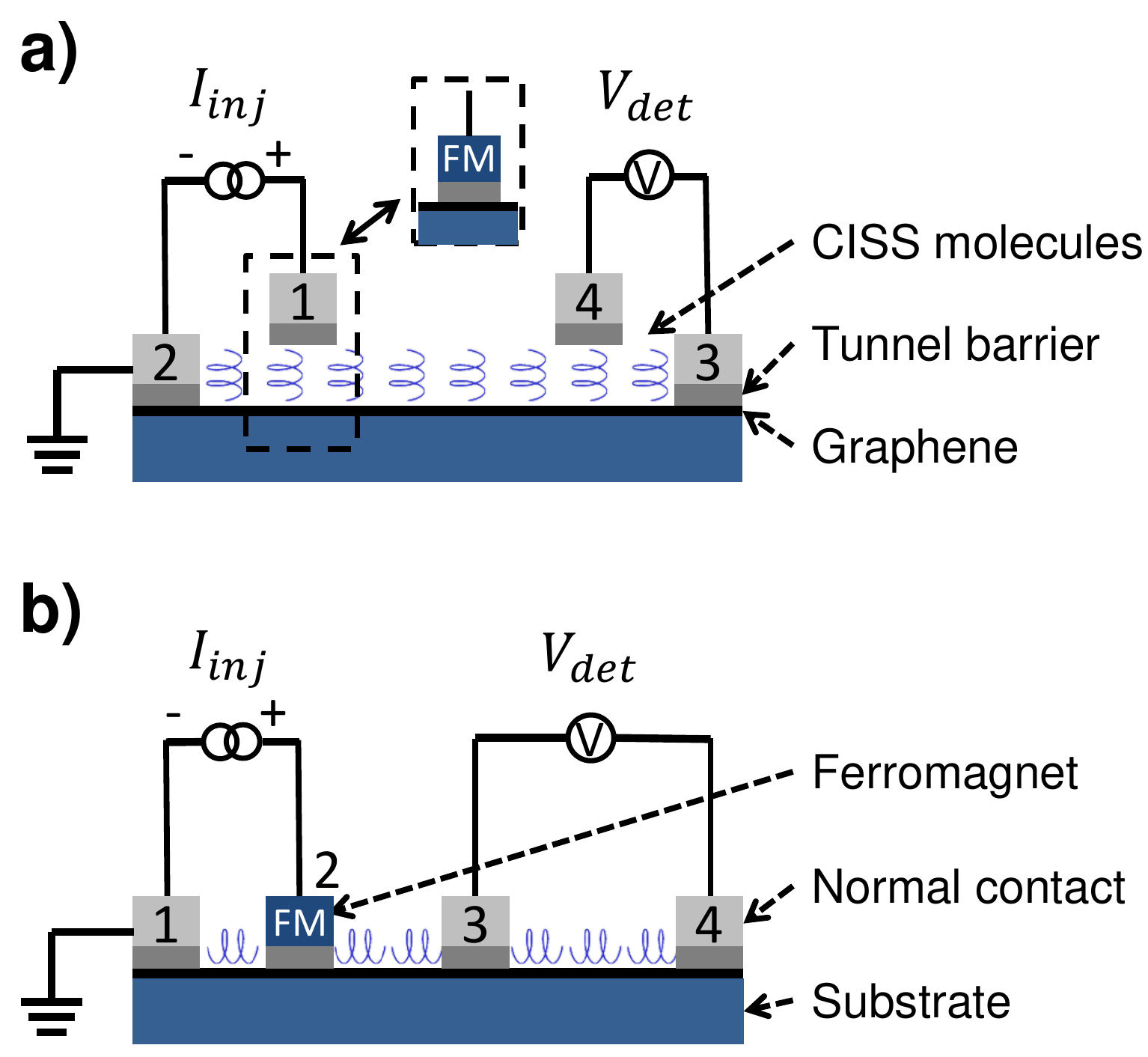}
	\caption{Nonlocal device designs with CISS molecules adsorbed on graphene. \textbf{a).} A device where electrons travel through CISS molecules. All contacts are non-magnetic and are numbered in agreement with Fig.~\ref{fig:4Tcross}\textbf{a)}. A variation of this device can be achieved by replacing contact $1$ with a ferromagnet, see inset. \textbf{b).} A device where electrons travel in proximity to CISS molecules. A ferromagnetic contact $2$ is used for spin injection, but one can also use only non-magnetic contacts, as in Fig.~\ref{fig:4Tlinear}.}\label{fig:nonlocal}
\end{figure}

Four-terminal measurements allow one to completely separate spin-related signals from charge-related signals, and therefore allow the detection of spin accumulations created by the CISS effect.\cite{jedema2001electrical} Here we analyze two geometries that are relevant for such measurements. In the first geometry, we use a node connected to BCB molecules to illustrate how spin injection and detection can occur without using magnetic materials (Fig.~\ref{fig:4Tcross}). In the second geometry, we use two nodes to decouple a BCB molecule from electrical contacts and illustrate the spin-charge conversion property of the molecule (Fig.~\ref{fig:4Tlinear}). In addition, we propose device designs that resemble these two geometries and discuss possible experimental outcomes (Fig.~\ref{fig:nonlocal}).

Fig.~\ref{fig:4Tcross}\textbf{a)} shows a geometry where a node is connected to four contacts. Two of the contacts contain BCB molecules, and the other two contain non-magnetic (tunnel) barriers. We consider an experiment where contacts $1$ and $2$ are used for current injection and contacts $3$ and $4$ are used for voltage detection. In terms of spin injection, we first assume that the voltage contacts $3$ and $4$ are weakly coupled to the node, and do not contribute to the spin accumulation in the node. This means that the chemical potentials of contacts $1$ and $2$ fully determine the spin-dependent chemical potential (\textcolor{black}{column} vector) of the node \textcolor{black}{$\boldsymbol{\mu_{node}}=\begin{pmatrix} \mu_{node \rightarrow}\\\mu_{node \leftarrow}\end{pmatrix}$}. We also assume $\mu_2=0$ for convenience since only the chemical potential difference between the two contacts is relevant. Under these assumptions, the node receives electrons only from contact $1$, but emits electrons to both contact $1$ and $2$. Therefore, the incoming current \textcolor{black}{(column vector)} into the node is
\begin{equation}
\color{black}
\boldsymbol{I_{in}}=\frac{G}{e} \mathbb{T}^P_R \mu_1 \begin{pmatrix}
1\\1
\end{pmatrix}  , 
\end{equation}
and the outgoing current \textcolor{black}{(column vector)} from the node is
\begin{equation}
\color{black}
\boldsymbol{I_{out}}=\frac{G}{e}\Big((\mathbb{I}-\mathbb{R}^P_L) + (1-r_{B})\mathbb{I}\Big) \boldsymbol{\mu_{node}},
\end{equation}
where $G=Ne^2/h$ is the $N$-channel, one-spin conductance of the channels connecting the node to each of the contacts, and $r_B$ ($0\leq r_B\leq1$) is the reflection probability of the tunnel barrier between the node and contact $2$ (different from the barriers in BCB molecules). Due to the spin-preserving nature of the node, at steady state the incoming current is equal to the outgoing current (for both spin components), or $\boldsymbol{I_{in}}=\boldsymbol{I_{out}}$. From this relation we derive
\begin{equation}
\color{black}
\boldsymbol{\mu_{node}}=\Big((1+t_B)\mathbb{I}-\mathbb{R}^P_L\Big)^{-1} \mathbb{T}^P_R \mu_1 \begin{pmatrix}
1\\1
\end{pmatrix},
\end{equation}
where $t_B=1-r_B$ is the transmission probability of the tunnel barrier. Next, we derive the spin accumulation in the node 
\begin{equation}\label{eqn:mus}
\mu_s=\mu_{node \rightarrow}-\mu_{node \leftarrow}\textcolor{black}{=\Big(1,-1\Big)\boldsymbol{\mu_{node}}}=k_{inj}\mu_1,
\end{equation}
where \textcolor{black}{a row vector $\Big(1,-1\Big)$ is used as an operator to calculate the difference between the two spin chemical potentials, and}
\begin{equation}
\color{black}
\begin{split}
k_{inj}=\Big(1,-1\Big) \Big((1+&t_B)\mathbb{I}-\mathbb{R}^P_L\Big)^{-1} \mathbb{T}^P_R \begin{pmatrix}
1\\1
\end{pmatrix},\\
\text{~with~} 0&< k_{inj} \leq \frac{1}{4},
\end{split}
\end{equation}
is the spin \textit{injection coefficient} for these current contacts. This expression shows that the spin accumulation in the node depends linearly on the chemical potential difference between the current contacts, and the coefficient $k_{inj}$ is determined by both the BCB molecule (with parameter $t$) and the tunnel barrier connected to contact $2$ (with parameter $t_B$).  

With regard to spin detection, we discuss whether the established spin accumulation $\mu_s$ in the node can lead to a chemical potential difference (and thus a charge voltage) between the weakly coupled voltage contacts $3$ and $4$. A contact cannot distinguish between the two spin components, therefore only the charge current (sum of both spins, \textcolor{black}{calculated by applying an operator $\Big(1,1\Big)$ to a current column vector}) is relevant. At steady state, there is no net charge current at any of the voltage contacts,
\begin{subequations}
	\color{black}
	\begin{equation}
	I_3=\frac{G}{e} \Big(1,1\Big) \Bigg((1-r_B) \mu_3 \begin{pmatrix}
	1\\
	1
	\end{pmatrix} -t_B \boldsymbol{\mu_{node}} \Bigg)=0,
	\end{equation}
	\begin{equation}
	I_4=\frac{G}{e} \Big(1,1\Big) \Bigg((\mathbb{I}-\mathbb{R}^P_L) \mu_4 \begin{pmatrix}
	1\\
	1
	\end{pmatrix} -\mathbb{T}^P_R \boldsymbol{\mu_{node}} \Bigg) =0,
\end{equation}
\end{subequations}
which gives
\begin{equation} \label{eqn:v34}
\mu_4-\mu_3=k_{det}\mu_s,
\end{equation}
where
\begin{equation}
\color{black}
\begin{split}
k_{det}=\frac{1}{2} &\frac{\begin{pmatrix}
	1,1
	\end{pmatrix}\mathbb{T}^P_R\begin{pmatrix}
	1\\-1
	\end{pmatrix}}{\begin{pmatrix}
	1,1
	\end{pmatrix}\mathbb{T}^P_L\begin{pmatrix}
	1\\1
	\end{pmatrix}},\\
\text{~with~}&0< k_{det} \leq \frac{1}{2},
\end{split}
\end{equation}
is the spin \textit{detection coefficient} for these voltage contacts. This expression shows that the chemical potential difference between the two voltage contacts depends linearly on the spin accumulation in the node, and the coefficient $k_{det}$ is exclusively determined by the BCB molecule (with parameter $t$).

Combining Eqn.~\ref{eqn:mus} and Eqn.~\ref{eqn:v34} we obtain
\begin{equation}
\frac{R_{4T}}{R_{2T}}=\frac{\mu_4-\mu_3}{\mu_1-\mu_2}=k_{inj}k_{det},
\end{equation}
where $R_{4T}$ is the four-terminal resistance (measured using contacts $3$ and $4$ as voltage contacts, while using contacts $1$ and $2$ as current contacts), and $R_{2T}$ is the two-terminal resistance (measured using contacts $1$ and $2$ as both voltage and current contacts). This ratio is determined by both the BCB molecule (with parameter $t$) and the tunnel barrier connected to contact $2$ (with parameter $t_B$), and can be experimentally measured to quantitatively characterize the CISS effect.

As an example, for $t=t_B=0.5$, we have $k_{inj}\approx0.11$, $k_{det}\approx0.17$, and $R_{4T}/R_{2T}\approx0.02 $. In Fig.~\ref{fig:4Tcross}\textbf{b)} we plot $R_{4T}/R_{2T}$ as a function of $t$ for three different $t_B$ values. Similar plots for $k_{inj}$ and $k_{det}$ are shown in Appendix~\ref{app:BCB}.

The above results show that it is possible to inject and detect a spin accumulation in a node using only BCB molecules and non-magnetic (tunnel) barriers, and these processes can be quantitatively described by the injection and detection coefficients. We stress that the signs of the injection and detection coefficients depend on the type (chirality) of the BCB molecule and the position of the molecule with respect to the contact. Switching the molecule from \textit{P}-type to \textit{AP}-type leads to a sign change of the injection or detection coefficient. The sign change also happens if the contact is connected to the opposite side of the BCB molecule. \textcolor{black}{For example, in Fig.~\ref{fig:4Tcross}\textbf{a)}, contacts $1$ and $4$ are both connected to the node via \textit{P}-type BCB molecules, but contact $1$ is on the left-hand side of a molecule, while contact $4$ is on the right-hand side. Electrons emitted from these two contacts travel in opposite directions through the (same type of) BCB molecules before arriving at the node. As a result, using contact $4$ instead of contact $1$ as a current contact leads to a sign change of $k_{inj}$. Similarly, using contact $1$ instead of contact $4$ as a voltage contact leads to a sign change of $k_{det}$. Experimentally, one can use three BCB contacts to observe this sign change: A fixed current contact (thus a fixed $k_{inj}$) in combination with two voltage contacts that use the same type of BCB molecule but are placed on opposite sides of a node (thus opposite signs for $k_{det}$). The voltages measured by the two voltage contacts (with respect to a common reference contact) will differ by sign.} This can be experimentally measured as a signature of the CISS effect.

Fig.~\ref{fig:4Tlinear} shows a geometry where a BCB molecule is between two nodes $A$ and $B$, and is decoupled from the contacts. The nodes themselves are connected to contacts in a similar fashion as in the previous geometry. In node $A$, we consider a chemical potential vector $\boldsymbol{\mu_{A}}$ and a spin accumulation $\mu_{sA}$, which are fully determined by the current contacts $1$ and $2$. In node $B$, we consider weakly coupled voltage contacts $3$ and $4$, so that its chemical potential vector $\boldsymbol{\mu_{B}}$ and its spin accumulation $\mu_{sB}$, are fully determined by $\boldsymbol{\mu_{A}}$. At steady state, there is no net charge or spin current in node $B$, which leads to 
\begin{equation}\label{eqn:scconv}
\color{black}
\boldsymbol{\mu_{B}}=(\mathbb{I}-\mathbb{R}^P_L)^{-1} \mathbb{T}^P_R \boldsymbol{\mu_{A}}.
\end{equation}
Note that here the matrices only refer to the molecule between the two nodes. For BCB molecules, this expression always gives $\mu_{sB}=0$, but for a more generalized CISS molecule (as described in Appendix~\ref{app:general}), this expression can give $\mu_{sB}\neq0$. This shows that a spin accumulation at one side of a CISS molecule can generate a spin accumulation at the other side of the molecule. Most importantly, for both the BCB model and the more generalized model, Eqn.~\ref{eqn:scconv} predicts that a spin accumulation difference across a CISS molecule creates a charge voltage across the molecule, and \textit{vice versa} (spin-charge conversion via a CISS molecule). Mathematically written, the expression always provides $\mu_{nA}\neq\mu_{nB}$ when $\mu_{sA}\neq \mu_{sB}$, and $\mu_{sA}\neq \mu_{sB}$ when $\mu_{nA}\neq\mu_{nB}$, where $\mu_{nA}$ (or $\mu_{nB}$) is the average chemical potential of the two spin components in node $A$ (or in node $B$). A more detailed description of this geometry can be found in Appendix~\ref{app:general}.

Fig.~\ref{fig:nonlocal} shows two types of nonlocal devices that resemble the two geometries introduced above. We realize the node function with graphene, chosen for its long spin lifetime and long spin diffusion length.\cite{drogeler2016spin} 

The first type, as shown in Fig.~\ref{fig:nonlocal}\textbf{a)}, represents the geometry in Fig.~\ref{fig:4Tcross}\textbf{a)}, where spin injection and detection are both achieved using CISS molecules. A current $I_{inj}$ is injected from contact $1$ through CISS molecules into graphene, then driven out to a normal metal contact $2$. This current induces a spin accumulation in the graphene layer underneath the current contacts, which then diffuses to the voltage contacts. The voltage contacts then pick up a charge voltage $V_{det}$ in a similar fashion as explained in Fig.~\ref{fig:4Tcross}\textbf{a)}. With this, the nonlocal resistance can be determined $R_{nl}=V_{det}/I_{inj}$. Further, we can derive (see Appendix~\ref{app:BCB} for details)
\begin{equation}\label{eqn:rnl1} \color{black}
R_{nl}=- k_{inj} k_{det} R_{inj} e^{-\frac{d}{\lambda_s}},
\end{equation}
where $R_{inj}$ is the resistance measured between the current contacts $1$ and $2$, $d$ is the distance between contacts $1$ and $4$, and $\lambda_s$ is the spin diffusion length of graphene. It is assumed here that the spacing between the current contacts ($1$ and $2$) and the spacing between the voltage contacts ($3$ and $4$) are both much smaller than $\lambda_s$. The minus sign comes from the fact that the injection and the detection contacts are on the same side of the graphene channel (both on top), unlike the example in Fig.~\ref{fig:4Tcross}\textbf{a)} (one on the left and the other on the right). 

A variation of this device is obtained by replacing contact $1$ (together with the CISS molecules underneath it) with a ferromagnet, as shown in the inset of Fig.~\ref{fig:nonlocal}\textbf{a)}. This variation allows one to control the sign of $R_{nl}$ by controlling the magnetization direction of the ferromagnet, which should be aligned parallel or anti-parallel to the helical (chiral) axis of the CISS molecules (out-of-plane, as indicated by the arrows). The nonlocal resistance is therefore
\begin{equation}\label{eqn:rnl2} \color{black}
R_{nl}(\Uparrow)=-R_{nl}(\Downarrow)= P_{FM} k_{det} R_{\lambda} e^{-\frac{d}{\lambda_s}},
\end{equation}
where the arrows indicate the magnetization directions, $P_{FM}$ is the polarization of the ferromagnet, and $R_{\lambda}$ is the spin resistance of graphene (see Appendix~\ref{app:BCB} for more details). In this device, the reversal of the magnetization direction of the ferromagnet leads to a sign change of the nonlocal resistance. \textcolor{black}{Under} experimental conditions,\cite{ingla201524,gurram2018spin} this nonlocal resistance change $\Delta R_{nl}= R_{nl}(\Uparrow)-R_{nl}(\Downarrow)$ can reach tens of Ohms ($\Omega$) and is easily detectable.

The second type of device is depicted in Panel \textbf{b)}. It is a variation of the geometry in Fig.~\ref{fig:4Tlinear}, where contact $2$ is replaced by a ferromagnet. In this device, instead of traveling through the CISS molecules, the electrons travel through the graphene channel underneath the molecules. It is assumed that due to the proximity of the CISS molecules, the electrons in graphene also experience a (weaker) CISS effect. Whether this assumption is valid remains to be proven. The nonlocal signals produced by this device are derived in Appendix~\ref{app:BCB}.

\section{Conclusion \label{con}}

In summary, we demonstrated that a spin-flip electron reflection process is inherent to the chiral induced spin selectivity (CISS) effect in linear-regime electron transport. Furthermore, we developed a set of spin-dependent electron transmission and reflection matrices and a generalized Barrier-CISS Center-Barrier (BCB) model to quantitatively describe the CISS effect in mesoscopic devices. Based on this formalism, we demonstrated that more than two terminals are needed in order to probe the CISS effect in linear-regime transport experiments. Moreover, we also showed several ways of injecting and detecting spins using CISS molecules and demonstrated that CISS molecules can give rise to spin-charge conversion. In addition, we proposed two types of graphene-based nonlocal devices which can be used to directly measure the CISS effect in the linear regime. 

We stress again that the above discussions and proposed devices are all based on linear-regime electron transport. Therefore, our conclusions cannot exclude the two-terminal detection of the CISS effect in the non-linear regime. However, the spin signals in non-linear-regime measurements should approach zero as the two-terminal bias approaches zero (entering the linear regime), and the mechanism that may contribute to such signals has to be different from spin-dependent electron transmission and reflection. A recent work shows that the CISS effect in electron photoemission experiments (three-terminal) can be explained by losses due to spin-dependent electron absorption in chiral molecules,\cite{nurenberg2018spin} but whether a similar process can lead to the detection of the CISS effect in non-linear two-terminal measurements remains to be investigated. \textcolor{black}{In general}, our model captures the fundamental role of the CISS effect in linear-regime mesoscopic devices \textcolor{black}{without assuming any microscopic electron transport mechanism inside CISS molecules.} Recently, a new type of spin-orbit coupling was predicted for one-dimensional screw dislocations in semiconductor crystals, which has a one-dimensional helical effective electric field. This type of spin-orbit coupling can lead to an enhanced spin lifetime for electrons traveling along the helical axis.\cite{hu2018ubiquitous} Future theoretical work should study whether similar effects exist in chiral or helical molecules.

In general, our model helps to analyze and understand device-based CISS experiments without having to understand the CISS effect on a molecular level. It provides a guideline for future reviewing and designing of CISS-based mesoscopic spintronic devices.

\section{Acknowledgments}

The authors acknowledge the financial support from the Zernike Institute for Advanced Materials (ZIAM), and the Spinoza prize awarded to Prof. B. J. van Wees by the Nederlandse Organisatie voor Wetenschappelijk Onderzoek (NWO). XY acknowledges helpful discussions with G.~E.~W.~Bauer, V.~Mujica, H.~Zacharias, and R.~Naaman.

\appendix
\section{General validity of the spin-flip reflection process}\label{app:sprefl}

In the main text, we stated that a spin-flip electron reflection process has to exist in order for spin-dependent transmission through CISS molecules to be allowed by the reciprocity theorem. Mathematically, this statements means that at least one off-diagonal term in the reflection matrices of CISS molecules has to be non-zero. Now we prove the general validity of our statement by limiting all off-diagonal terms of the reflection matrices to zero, and derive violations against the description of the CISS effect. \textcolor{black}{Under this limit, the general form of the transmission and reflection matrices of a CISS molecule is 
\begin{equation}
	\mathbb{T}_{CISS}=\begin{pmatrix}
	a&c\\b&d
	\end{pmatrix},~\mathbb{R}_{CISS}=\begin{pmatrix}
	A&0\\0&D
	\end{pmatrix},
\end{equation}
where $0 \leq a,b,c,d,A,D \leq 1$}. For electrons traveling towards the CISS molecule, each spin component can either be transmitted (with or without spin-flip) or reflected, the sum of these probabilities is therefore unity.
\begin{equation}
\color{black}
a+b+A=1,~c+d+D=1.	
\end{equation}
\textcolor{black}{Therefore we have}
\begin{equation}
	\mathbb{R}_{CISS}=\begin{pmatrix}
	1-a-b&0\\0&1-c-d
	\end{pmatrix}
\end{equation}
In addition, we adopt transmission and reflection matrices of a ferromagnet from Fig.~\ref{fig:matrices}.

Next, we consider that the CISS effect exists in the linear regime. This means that (according to the description of the CISS effect) with an input of spin non-polarized electrons, the CISS molecule gives a spin-polarized transmission output (a non-zero spin current $I_s$). Here we do not make assumptions about the chirality of the molecule or the electron flow direction, so that our conclusions hold for the most general situations. Therefore, we do not assume the sign of $I_s$, and write
\begin{equation}\label{seqn:ciss}
\color{black}
I_s=\Big( 1,-1 \Big) \mathbb{T}_{CISS} \begin{pmatrix}
1/2\\1/2
\end{pmatrix}=\frac{(a+c)-(b+d)}{2}\neq 0,
\end{equation}
where the \textcolor{black}{column} vector indicates the spin non-polarized input current, and the \textcolor{black}{row} vector is an operator that calculates the difference between the two spin components in the output current. 

In order to illustrate the discrepancy between the assumption of not having any spin-flip reflection and the conceptual description of the CISS effect (Eqn.~\ref{seqn:ciss}), we apply the reciprocity theorem to the circuit segment shown in Fig.~\ref{fig:2Tcfm}. For this circuit segment, we calculate the total transmission matrix accounting for the contribution from both the CISS molecule and the ferromagnet, and obtain
\begin{equation}
\color{black}
\mathbb{T}_{21}=\mathbb{T}_{FM} \cdot \Bigg(\mathbb{I}-\mathbb{R}_{CISS}\cdot\mathbb{R}_{FM} \Bigg)^{-1}\cdot \mathbb{T}_{CISS}.
\end{equation}
The transmission probability accounting for both spin species is therefore
\begin{equation}
T_{21}=\Big(1,1\Big)\mathbb{T}_{21}\begin{pmatrix}
1/2\\1/2
\end{pmatrix},
\end{equation}
\textcolor{black}{where the row vector is an operator that calculates the sum of both spin species. By substituting the matrices, we can write $T_{21}$ as a function of $P_{FM}$
\begin{equation}
\begin{split}
T_{21}(P_{FM})=&~\frac{a(1+P_{FM})}{1+a+b+P_{FM}(1-a-b)}\\
&+\frac{d(1-P_{FM})}{1+c+d-P_{FM}(1-c-d)}.
\end{split}
\end{equation}
Here the magnetization reversal of the ferromagnet is equivalent to a sign change of $P_{FM}$. Therefore,} the broader reciprocity theorem requires \cite{buttiker1988symmetry,buttiker1986four}
\begin{equation}
\color{black}
T_{12}(P_{FM})\equiv T_{12}(-P_{FM})
\end{equation}
for all $0<|P_{FM}|\leq1$. This requirement gives
$$
a=d, \text{and~} b=c,
$$
and therefore,
$$
(a+c)-(b+d)=0,
$$
which violates Eqn.~\ref{seqn:ciss}.

Therefore, we proved that the spin-flip electron reflection process has to exist in order for the CISS effect to exist in the linear transport regime, and this is a direct requirement from the \textit{Reciprocity Theorem}.

\section{Further discussions about the BCB model}\label{app:BCB}

\subsection{Derivation of the BCB transmission and reflection matrices}

In the BCB model, we consider a CISS molecule as two normal barriers sandwiching an ideal CISS center. The three parts together determine the transmission and reflection matrices of the molecule. We derive these matrices in two steps. First, we calculate the combined contribution of the left-most barrier and the ideal CISS center as Part~$1$ (superscript $P1$). Then, we add the second barrier to Part~1 and calculate the total effect.

We adopt the transmission and reflection matrices for a normal barrier and a CISS center from Fig.~\ref{fig:matrices}. Here we only discuss \textit{P}-type CISS molecules, as the \textit{AP}-type can be derived using the relations given in Fig.~\ref{fig:matrices}. For Part~$1$ we have
\begin{subequations}
\begin{equation}
\color{black}
\begin{split}
\mathbb{T}_{R}^{P1}&=(\mathbb{T}_{0,R}^P+  \mathbb{T}_{0,R}^P \cdot \mathbb{R}_B \cdot \mathbb{R}_{0,R}^P) \cdot \mathbb{T}_B\\
&=\begin{pmatrix}
t&t(1-t)\\0&0
\end{pmatrix},
\end{split}
\end{equation}
\begin{equation}
\color{black}
\begin{split}
\mathbb{R}_{R}^{P1}&=\mathbb{R}_B+\mathbb{T}_B \cdot \mathbb{R}_{0,R}^P \cdot \mathbb{T}_B\\&=\begin{pmatrix}
1-t&t^2\\0&1-t
\end{pmatrix},
\end{split}
\end{equation}
\begin{equation}
\color{black}
\begin{split}
\mathbb{T}_{L}^{P1}&=(\mathbb{T}_B + \mathbb{T}_B \cdot\mathbb{R}_{0,R}^P \cdot \mathbb{R}_B) \cdot \mathbb{T}_{0,L}^P\\&=\begin{pmatrix}
0&t(1-t)\\0&t
\end{pmatrix},
\end{split}
\end{equation}
\begin{equation}
\color{black}
\begin{split}
\mathbb{R}_{L}^{P1}&=\mathbb{R}_{0,L}^P+\mathbb{T}_{0,R}^P\cdot\mathbb{R}_B \cdot \mathbb{R}_{0,R}^P \cdot\mathbb{R}_B \cdot\mathbb{T}_{0,L}^P\\&=\begin{pmatrix}
0&(1-t)^2\\1&0
\end{pmatrix}.
\end{split}
\end{equation}
\end{subequations}
Note that here we have a finite number of reflections between the first barrier and the CISS center because the CISS center is ideal. Now we consider Part~1 as one unit and combine it with the second normal barrier
\begin{subequations}
\color{black}
\begin{equation}
\begin{split}
\mathbb{T}_R^P&=\mathbb{T}_B\cdot(\mathbb{I}- \mathbb{R}_{L}^{P1} \cdot \mathbb{R}_B)^{-1} \cdot \mathbb{T}_{R}^{P1} \\
&=\begin{pmatrix}
\sigma/(1-t) & \sigma \\ \sigma & \sigma (1-t)
\end{pmatrix},
\end{split}
\end{equation}
\begin{equation}
\begin{split}
\mathbb{R}_R^P&=\mathbb{R}_R^{P1}+\mathbb{T}_L^{P1}\cdot
(\mathbb{I}-\mathbb{R}_B \cdot \mathbb{R}_{L}^{P1})^{-1}\cdot\mathbb{R}_B\cdot\mathbb{T}_R^{P1}\\
&=\begin{pmatrix}
\sigma(1-t)^2-t+1 & \sigma/(1-t) \\ \sigma(1-t) & \sigma(1-t)^2-t+1)
\end{pmatrix},
\end{split}
\end{equation}
\begin{equation}
\begin{split}
\mathbb{T}_L^P&=\mathbb{T}_{L}^{P1} \cdot (\mathbb{I}-\mathbb{R}_B \cdot \mathbb{R}_{L}^{P1})^{-1} \cdot \mathbb{T}_B  \\
&=\begin{pmatrix}
\sigma(1-t) & \sigma \\ \sigma & \sigma /(1-t)
\end{pmatrix},
\end{split}
\end{equation}
\begin{equation}
\begin{split}
\mathbb{R}_L^P&=\mathbb{R}_B+\mathbb{T}_B \cdot(\mathbb{I}- \mathbb{R}_{L}^{P1} \cdot \mathbb{R}_B )^{-1} \cdot \mathbb{R}_L^{P1} \cdot \mathbb{T}_B\\
&=\begin{pmatrix}
\sigma(1-t)^2-t+1 & \sigma(1-t) \\ \sigma/(1-t) & \sigma(1-t)^2-t+1)
\end{pmatrix},
\end{split}
\end{equation}
\begin{equation*}
\text{where~} \sigma=\frac{t^2(1-t)}{-t^4+4t^3-6t^2+4t}.
\end{equation*}
\end{subequations}

These results show that in the BCB model, one parameter $t$ ($0<t\leq1$) determines the entire set of transmission and reflection probability matrices. Therefore it is possible to plot various transmission or reflection analysis results as functions of $t$, as shown in the supplementary information. With different $t$ values the BCB model is able to represent a large spectrum of CISS molecules with different "strengths" of the CISS effect. While it is sufficient to illustrate the fundamental role of a CISS molecule in a solid-state device, it is still a simplified picture. We will introduce a more generalized model later in Appendix~\ref{app:general}.

\subsection{Discussion on the FM-BCB geometry}

\begin{figure}[htb]
	\includegraphics[width=\linewidth]{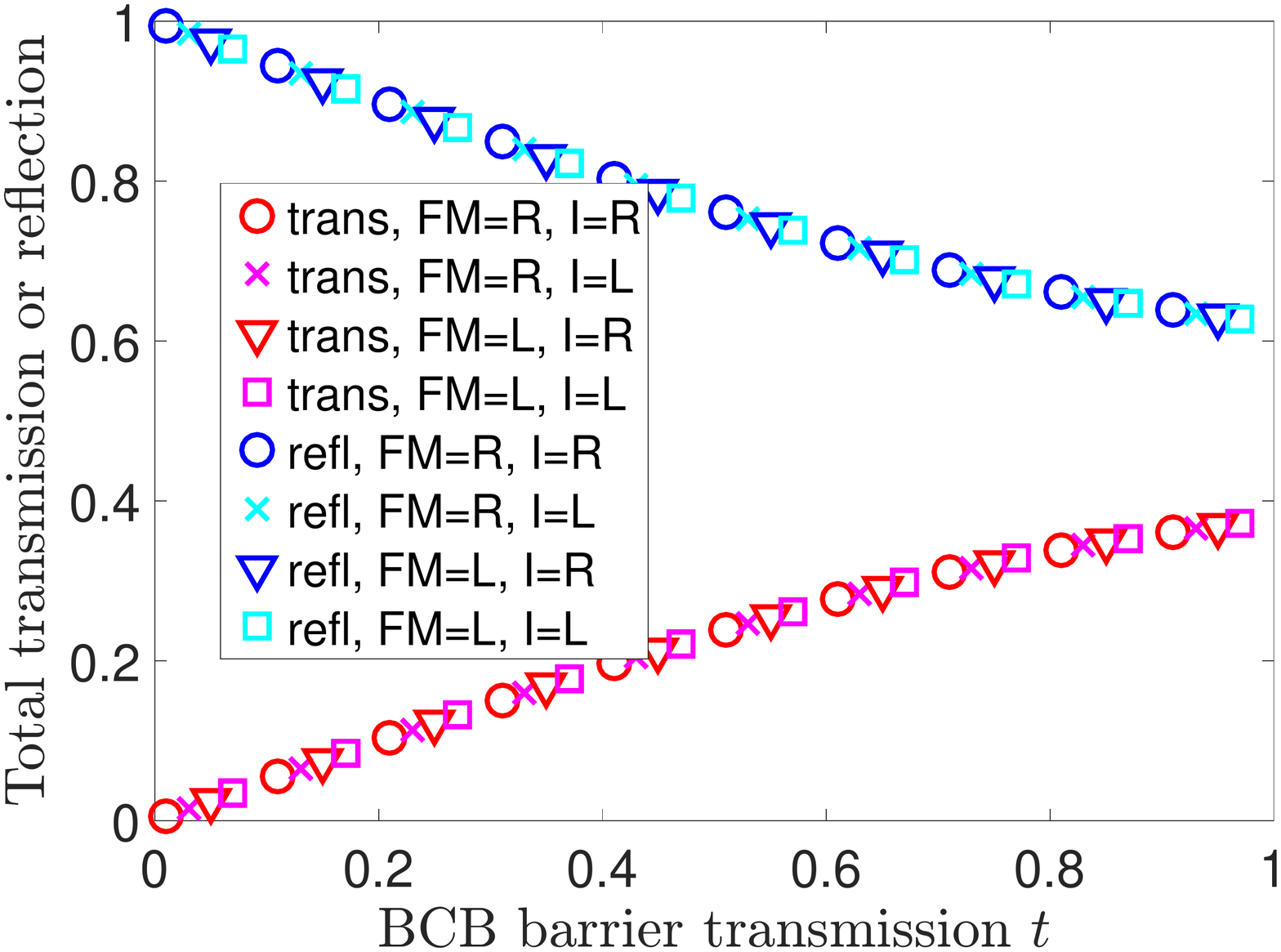}
	\caption{Normalized total transmission and reflection of an \textit{FM-BCB} segment as a function of $t$ (BCB barrier transmission). Red and magenta labels (lower group) are for total transmissions accounting for both spin species for different magnetization orientations (FM=R or L, for magnetization \textit{right} or \textit{left}) and different electron flow directions (I=R or L, for electron flow from \textit{left to right} or from \textit{right to left}). Blue and cyan labels (upper group) are for total reflections. The polarization of the ferromagnet is chosen as $P_{FM}=0.1$, comparable to experimental conditions with Co contacts.}
	\label{fig:t_FM_BCB}
\end{figure}

In the main text, we discussed that the two-terminal transmission remains constant under magnetization reversal in the \textit{FM-BCB} geometry, here we illustrate the same result under current reversal. \textcolor{black}{(Note that this is also a restriction from the linear regime.)} 

The transmission and reflection matrices from contact $2$ to contact $1$ are
\begin{subequations}
	\color{black}
	\begin{equation}
	\begin{split}
	&\mathbb{T}^{FM-BCB}_{12}(\Rightarrow)\\
	=& \mathbb{T}_{FM}(\Rightarrow) \cdot \Bigg(\mathbb{I}-\mathbb{R}^P_R \cdot \mathbb{R}_{FM}(\Rightarrow)  \Bigg)^{-1} \cdot \mathbb{T}^P_L,
	\end{split}
	\end{equation}
	\begin{equation}
	\begin{split}
	&\mathbb{R}^{FM-BCB}_{22}(\Rightarrow)\\
	=&\mathbb{R}^P_{L}+\mathbb{T}^P_R \cdot \Bigg(\mathbb{I}- \mathbb{R}_{FM}(\Rightarrow) \cdot \mathbb{R}^P_R\Bigg)^{-1}\\&~~~~ \cdot \mathbb{R}_{FM}(\Rightarrow) \cdot \mathbb{T}^P_{L},
	\end{split}
	\end{equation}
\end{subequations}
and the corresponding transmission and reflection probabilities accounting for both spins are
\begin{subequations}
	\color{black}
	\begin{equation}
	T^{FM-BCB}_{12}(\Rightarrow)=\Big(1,1\Big) \mathbb{T}^{FM-BCB}_{12}(\Rightarrow)\begin{pmatrix}
	1/2\\1/2 \end{pmatrix},
	\end{equation}
	\begin{equation}
	R^{FM-BCB}_{22}(\Rightarrow)=\Big(1,1\Big)\mathbb{R}^{FM-BCB}_{22}(\Rightarrow)\begin{pmatrix}	
	1/2\\1/2 \end{pmatrix}.
	\end{equation}
\end{subequations}
With these expressions, we can calculate the transmission and reflection probabilities as a function of $t$ (BCB transmission probability) for four situations: two current directions and two magnetization directions, and the results are plotted in Fig.~\ref{fig:t_FM_BCB}. Note that for all four situations, the transmission curves (or the reflection curves) completely overlap with each other, this means that neither magnetization reversal nor current reversal can lead to a signal change in the two-terminal conductance. Furthermore, we are able to quantitatively analyze the contribution of each spin component in each of the four situations, and this will be shown in the supplementary information.

\subsection{Discussion on the spin valve geometry}

\begin{figure}[htb]
	\includegraphics[width=\linewidth]{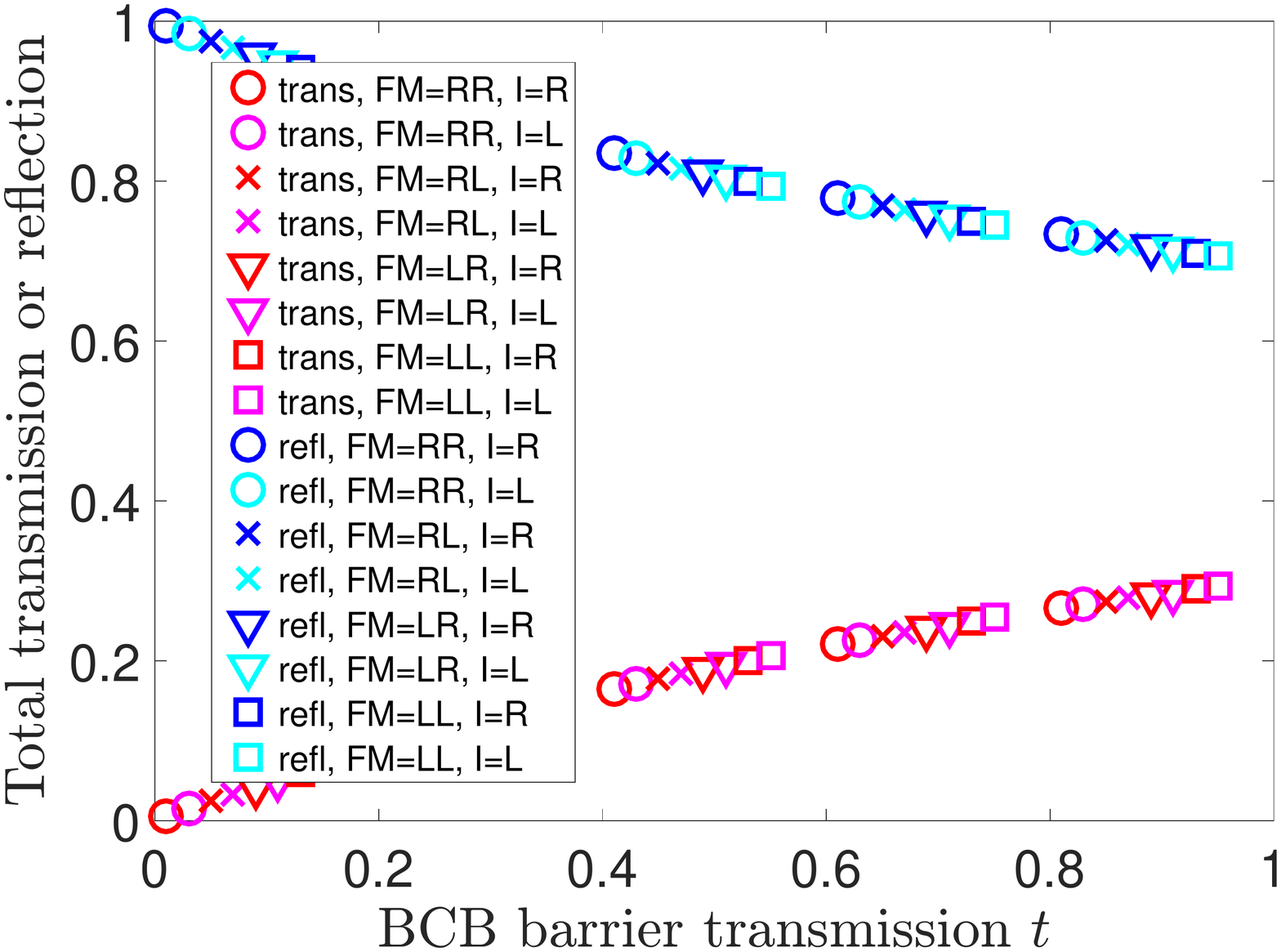}
	\caption{Normalized total transmission and reflection of a spin-valve segment as a function of $t$ (BCB barrier transmission). Red and magenta labels (lower group) are for total transmissions accounting for both spin species for different ferromagnet configurations (similar notation as before, but for two ferromagnets, FM=RR, RL, LR, or LL) and different current flow directions (I=R or L). Blue and cyan labels (upper group) are for total reflections. The polarization of the ferromagnet is chosen as $P_{FM}=0.1$, comparable to experimental conditions with Co contacts.}
	\label{fig:t_SV}
\end{figure}

Similar to the \textit{FM-BCB} geometry, in order to calculate the two-terminal transmission and reflection probability of the spin valve geometry, we first calculate the transmission and reflection matrices for this geometry. For this, we treat the spin valve geometry as an \textit{FM-BCB} module and a ferromagnet connected in series, and derive
\begin{subequations}
	\color{black}
\begin{equation}
\begin{split}
&\mathbb{T}^{SV}_{21}(\Rightarrow,\Rightarrow)\\
=& \mathbb{T}_{FM}(\Rightarrow) \cdot \Bigg(\mathbb{I}- \mathbb{R}^{FM-BCB}_{22}(\Rightarrow) \cdot \mathbb{R}_{FM}(\Rightarrow)\Bigg)^{-1}\\
&\cdot \mathbb{T}^{FM-BCB}_{21}(\Rightarrow), 
\end{split}
\end{equation}
\begin{equation}
\begin{split}
&\mathbb{R}^{SV}_{11}(\Rightarrow,\Rightarrow)\\
=&\mathbb{R}^{FM-BCB}_{11}(\Rightarrow)+\mathbb{T}^{FM-BCB}_{21}(\Rightarrow)  \\
&~~~~ \cdot \Bigg(\mathbb{I}- \mathbb{R}_{FM}(\Rightarrow) \cdot \mathbb{R}^{FM-BCB}_{22}(\Rightarrow) \Bigg)^{-1} \\
&~~~~ \cdot \mathbb{R}_{FM}(\Rightarrow) \cdot \mathbb{T}^{FM-BCB}_{12}(\Rightarrow),
\end{split}
\end{equation} 
\begin{equation}
\begin{split}
&\mathbb{T}^{SV}_{12}(\Rightarrow,\Rightarrow)\\
=&\mathbb{T}^{FM-BCB}_{12}(\Rightarrow) \cdot \Bigg(\mathbb{I}- \mathbb{R}_{FM}(\Rightarrow) \cdot \mathbb{R}^{FM-BCB}_{22}(\Rightarrow)\Bigg)^{-1} \\
&\cdot \mathbb{T}_{FM}(\Rightarrow),
\end{split}
\end{equation} 
\begin{equation}
\begin{split}
&\mathbb{R}^{SV}_{22}(\Rightarrow,\Rightarrow)\\
=&\mathbb{R}_{FM}(\Rightarrow)+\mathbb{T}_{FM}(\Rightarrow) \\ 
&~~~~\cdot \Bigg(\mathbb{I}- \mathbb{R}^{FM-BCB}_{22}(\Rightarrow) \cdot \mathbb{R}_{FM}(\Rightarrow) \Bigg)^{-1}\\
&~~~~\cdot \mathbb{R}^{FM-BCB}_{22}(\Rightarrow) \cdot \mathbb{T}_{FM}(\Rightarrow),
\end{split}
\end{equation} 
\end{subequations}
where the two arrows in the brackets on the left-hand side of the equations indicate the magnetization direction of the two ferromagnets respectively. For the case where the magnetization of one of the ferromagnets is reversed, we can substitute the corresponding magnetization direction with an opposite arrow. 

For the two-terminal transmission and reflection probabilities accounting for both spins, we have
\begin{subequations}
	\color{black}
	\begin{equation}
	T^{SV}_{ij}=\Big(1,1\Big) \cdot \mathbb{T}^{SV}_{ij} \cdot \begin{pmatrix}
	1/2\\
	1/2
	\end{pmatrix},
	\end{equation}
	\begin{equation}
	R^{SV}_{ij}=\Big(1,1\Big) \cdot \mathbb{R}^{SV}_{ij} \cdot \begin{pmatrix}
	1/2\\
	1/2
	\end{pmatrix},
	\end{equation}
\end{subequations}
and we can calculate these probabilities as a function of $t$ (BCB transmission probability) for eight situations: the two ferromagnets each with two magnetization directions, and two opposite current directions, as shown in Fig.~\ref{fig:t_SV}. Note that for all eight situations, the transmission curves (or the reflection curves) completely overlap with each other, this means that neither magnetization reversal (for either ferromagnet) nor current reversal can lead to a signal change in the two-terminal conductance. Furthermore, we can quantitatively analyze the contribution of each spin component in this geometry, as shown in the supplementary information.

\subsection{Discussion on injection and detection coefficients}

\begin{figure}[htb]
	\includegraphics[width=\linewidth]{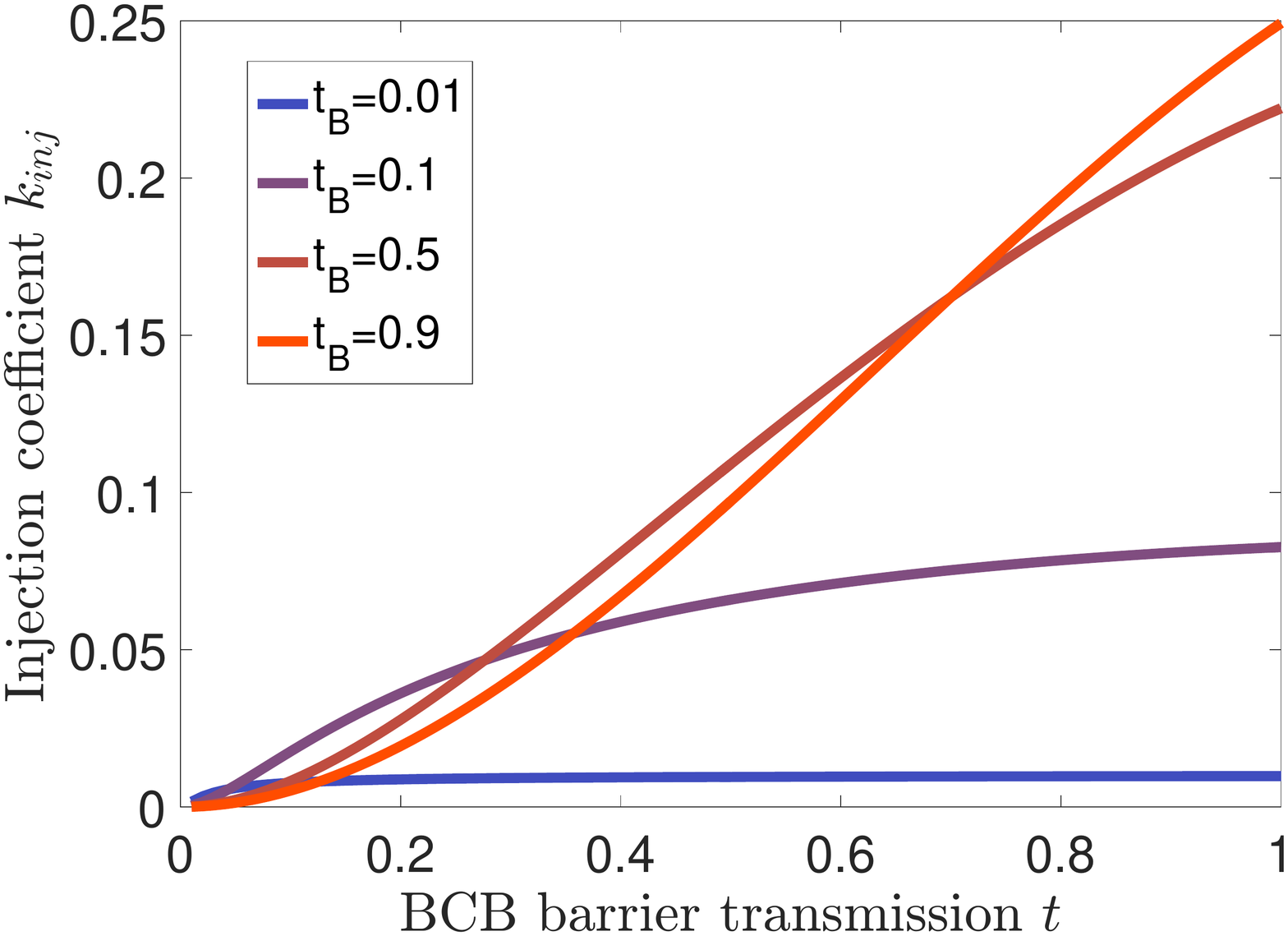}
	\caption{Injection coefficient $k_{inj}$ as a function of $t$ (BCB barrier transmission) for various $t_B$ (contact barrier transmissiion), for the geometry described in Fig.~\ref{fig:4Tcross}\textbf{a)} in the main text.}
	\label{fig:t_kinj}
\end{figure}

\begin{figure}[htb]
	\includegraphics[width=\linewidth]{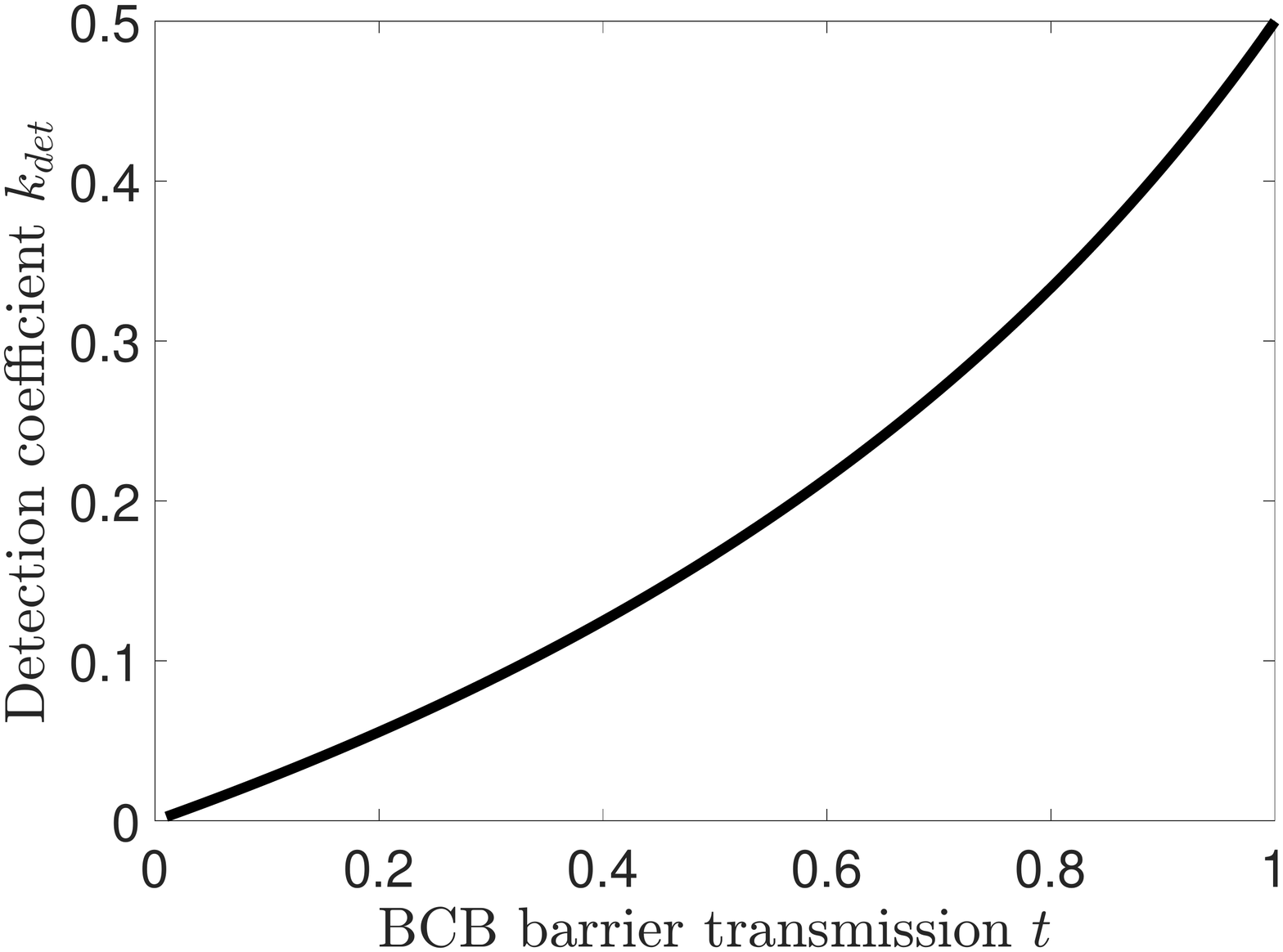}
	\caption{Detection coefficient $k_{det}$ as a function of $t$ (BCB barrier transmission) for the geometry described in Fig.~\ref{fig:4Tcross}\textbf{a)} in the main text.}
	\label{fig:t_kdet}
\end{figure}

\begin{figure}[htb]
	\includegraphics[width=\linewidth]{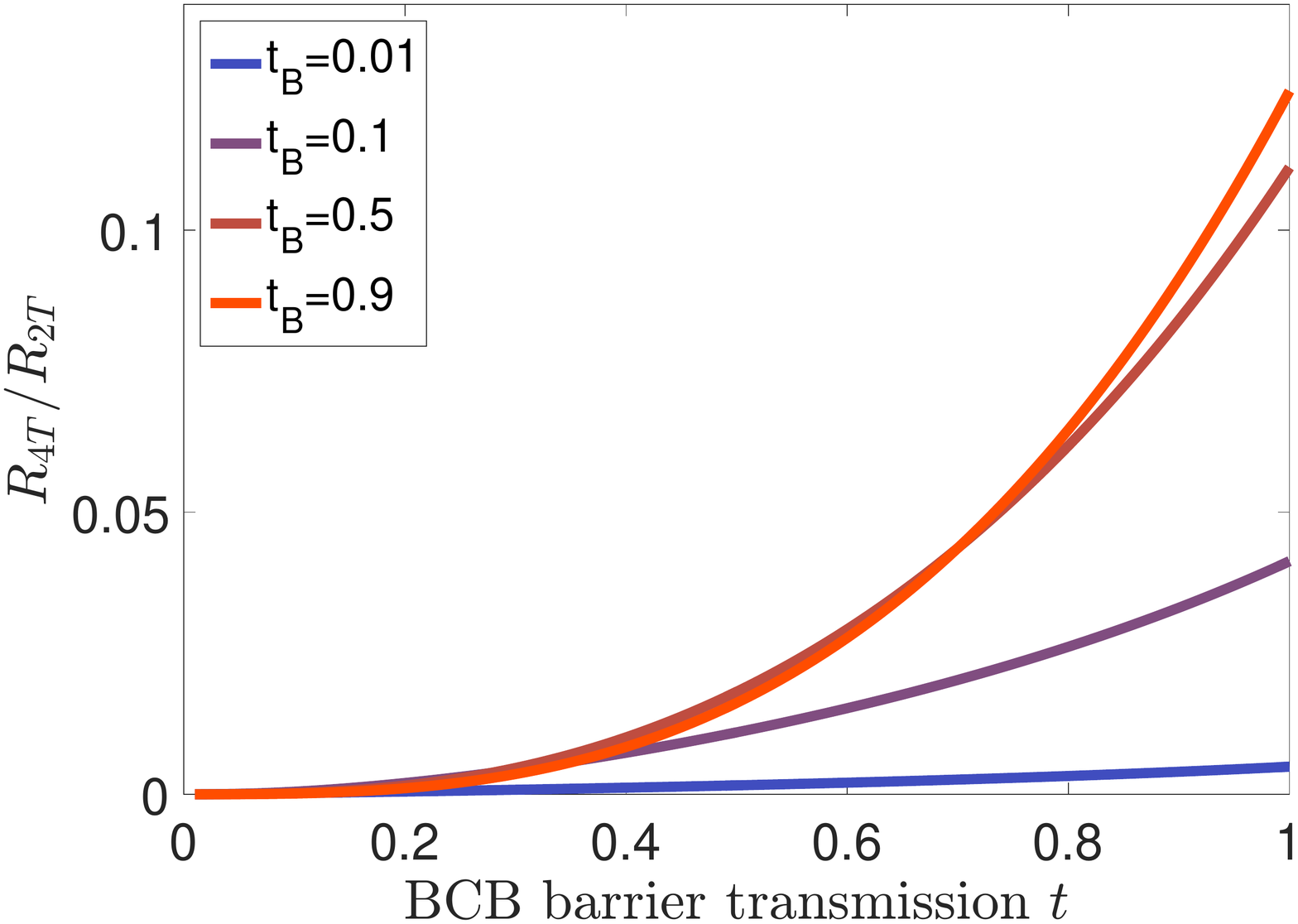}
	\caption{The ratio between four-terminal and two-terminal resistances as a function of $t$ (BCB barrier transmission) for various $t_B$ (contact barrier transmissiion), for the geometry described in Fig.~\ref{fig:4Tcross}\textbf{a)} in the main text.}
	\label{fig:t_4t2t}
\end{figure}

In the main text, we derived the injection and detection coefficients $k_{inj}$ and $k_{det}$ for the four-terminal geometry shown in Fig.~\ref{fig:4Tcross}\textbf{a)}, and we showed that the product of these two geometries represents the ratio between four-terminal and two-terminal resistances. The injection coefficient depends both on the BCB barrier transmission $t$ and the transmission probability $t_B$ of the barrier at contact $2$, whereas the detection coefficient only depends on the BCB barrier transmission $t$. This difference is due to our assumption of weakly coupled detection contacts. Here we plot the injection coefficient $k_{inj}$, the detection coefficient $k_{det}$, and the 4T/2T resistance ratio $R_{4T}/R_{2T}$ as a function of $t$ (transmission probability of the barrier in the BCB molecule). Especially, for $k_{inj}$ and $R_{4T}/R_{2T}$, we set $t_B$ (transmission probability of the barrier in contact $2$) to a few different values and illustrate its influence.

\subsection{Discussion on spin-charge conversion} 

In the main text, we illustrated the spin-charge conversion property of CISS molecules. The chemical potential vectors in the two nodes in Fig.~\ref{fig:4Tlinear} are related to each other following
Eqn.~\ref{eqn:scconv}. We also introduced the scalers of charge chemical potential $\mu_n$ (the average of two spin chemical potentials) and spin accumulation $\mu_s$ (the difference between two spin chemical potentials). Here we show how these scaler chemical potentials relate to each other.

A vector chemical potential of a node can be rewritten as a sum of two column vectors
\begin{equation}
\color{black}
\begin{split}
\boldsymbol{\mu}&=\begin{pmatrix}\mu_{\rightarrow}\\\mu_{\leftarrow}\end{pmatrix}=\begin{pmatrix}\mu_n+\frac{1}{2}\mu_s \\ \mu_n-\frac{1}{2}\mu_s
\end{pmatrix}\\
&=\mu_n \begin{pmatrix}
1\\1
\end{pmatrix} +\frac{1}{2}\mu_s \begin{pmatrix}
1\\-1
\end{pmatrix}.
\end{split}
\end{equation}
By rewriting both $\boldsymbol{\mu_A}$ and $\boldsymbol{\mu_B}$ in this fashion, and substituting them into Eqn.~\ref{eqn:scconv}, we obtain
\begin{equation}\label{seqn:muns}
\color{black}
\Delta \mu_n \begin{pmatrix}
1\\1
\end{pmatrix}+\frac{1}{2}\Delta \mu_s \begin{pmatrix}
1\\-1
\end{pmatrix}= \Bigg(\mathbb{I}-(\mathbb{I}-\mathbb{R}^P_L)^{-1}\mathbb{T}^P_R\Bigg) \boldsymbol{\mu_A},
\end{equation}
where $\Delta \mu_n=\mu_{nA}-\mu_{nB}$ is the charge chemical potential difference between the two nodes, and $\Delta \mu_s=\mu_{sA}-\mu_{sB}$ is the spin accumulation difference between the two nodes.

For the BCB model,
\begin{equation}
\color{black}
\mathbb{I}-(\mathbb{I}-\mathbb{R}^P_L)^{-1}\mathbb{T}^P_R=\begin{pmatrix}
1+\frac{1}{t-2}&-1-\frac{1}{t-2}\\
\frac{1}{t-2}&-\frac{1}{t-2}
\end{pmatrix}
\end{equation}
where $t$ is the BCB transmission probability. Substituting this matrix into Eqn.~\ref{seqn:muns} gives
\begin{equation}
\color{black}
\Delta \mu_n \begin{pmatrix}
1\\1
\end{pmatrix}+\frac{1}{2}\Delta \mu_s \begin{pmatrix}
1\\-1
\end{pmatrix}=\mu_{sA}\begin{pmatrix}
1+\frac{1}{t-2}\\ \frac{1}{t-2}
\end{pmatrix}.
\end{equation}
Note that here the charge chemical potentials in node $A$ ($\mu_{nA}$) drops out of the equation. We emphasize that this is a unique result for the BCB model. 

Solving the above vector equation gives
\begin{subequations}
	\begin{equation}\label{seqn:s6a}
	\Delta \mu_s= \mu_{sA},
	\end{equation}
	\begin{equation}\label{seqn:s6b}
	\Delta \mu_n = -k_{conv} \Delta \mu_s,
	\end{equation}
\end{subequations}
where $k_{conv}=t/(4-2t)$ ($0< k_{conv}\leq 0.5$) is the spin-charge \textit{conversion coefficient}. 

Eqn.~\ref{seqn:s6a} shows that $\mu_{sB}=\mu_{sA}-\Delta \mu_s=0$, meaning that a spin accumulation in node $A$ cannot generate a spin accumulation in node $B$. This result is special for the BCB model and does not hold for a more generalized CISS model (as will be introduced later in Appendix~\ref{app:general}). Eqn.~\ref{seqn:s6b} shows that the charge voltage across a BCB molecule linearly depends on the spin accumulation difference across the molecule, and \textit{vice versa} (spin-charge conversion). The conversion coefficient $k_{conv}$ depends on $t$ (BCB barrier transmission). Notably, for BCB molecules this coefficient has the same value as the detection coefficient ($k_{conv}=k_{det}$). This is because the result $\mu_{sB}=0$ makes node $B$ equivalent to a contact: it is not able to distinguish between two spin components. For a more generalized model (as will be introduced later in Appendix~\ref{app:general}) the two coefficient take different values. These conclusions were mentioned in the main text, here we showed the proof.

\subsection{Discussion on nonlocal signals}

We first discuss the non-local resistance measured with device geometries shown in Fig.~\ref{fig:nonlocal}\textbf{a)}. In this picture the axes of the CISS molecules are vertical rather than horizontal, therefore the spin orientations are described as spin-up or spin-down. As a result, the spin accumulation is defined as $\mu_{s}=(\mu_{\uparrow}-\mu_{\downarrow})$. 

First, we discuss the case where the spin injection is done through a BCB molecule. The injected spin accumulation underneath contact $1$ is
\begin{equation}
\mu_{s,inj}=-k_{inj}\mu_1=-e k_{inj} I_{inj} R_{inj},
\end{equation}
where the minus sign is due to the fact that the electrons are traveling downwards through CISS molecules into graphene, as a result, the injected spin accumulation is negative (mostly spin-down). The resistance \begin{equation}
R_{inj}=R_{12}=\frac{\mu_1-\mu_2}{eI_{inj}}
\end{equation}
is the resistance measured between the two injection contacts $1$ and $2$.

Inside graphene, the spin accumulation diffuses to all directions, and therefore the spin accumulation at the detection contact is
\begin{equation} \color{black}
\mu_{s,det}=\mu_{s,inj} e^{-\frac{d}{\lambda_s}},
\end{equation}
where $\lambda_s$ is the spin diffusion length in graphene, and $d$ is the distance between the inner injection and detection contacts ($1$ and $4$) (we have assumed that this distance is much larger than the separation of the two injection contacts or the separation of the two detection contacts). 

Further, the voltage detected by the detection contacts is (following Eqn.~\ref{eqn:v34})
\begin{equation}
V_{det}=\frac{1}{e}k_{det} \mu_{s,det}.
\end{equation}
With this, we have 
\begin{equation} \color{black}
R_{nl}=\frac{V_{det}}{I_{inj}}=-k_{inj}k_{det}R_{inj} e^{-\frac{d}{\lambda_s}},
\end{equation}
as in Eqn.~\ref{eqn:rnl1}.

For the case where the spin injection is obtained through a ferromagnet, the spin injection becomes
\begin{equation}
\mu_{s,inj}=\pm e P_{FM} I_{inj} R_{\lambda},
\end{equation}
where $P_{FM}$ is the polarization of the ferromagnet (with magnetization direction out-of-plane), and $R_{\lambda}$ is the spin resistance of graphene. This spin resistance is determined by the spin relaxation length in graphene and the shape of the graphene channel, and is defined as $R_{\lambda}=R_{sq}\lambda_{s}/W$, where $R_{sq}$ is the square resistance of graphene and $W$ is the width of the graphene channel (assuming the channel width remains the same across the spin diffusion length). The sign of the injected spin accumulation is determined by the magnetization direction of the ferromagnet, with magnetization-up for positive spin accumulation and magnetization-down for negative spin accumulation.

The diffusion and detection mechanisms are the same as in the previous case. Therefore, the non-local resistance for this situation is 
\begin{equation} \color{black}
R_{nl}=\pm k_{det} P_{FM} R_{\lambda} e^{-\frac{d}{\lambda_s}},
\end{equation} 
as in Eqn.~\ref{eqn:rnl2}. With the help of the ferromagnet, it is possible to switch the sign of the non-local resistance.

Next, for the device shown in Fig.~\ref{fig:nonlocal}\textbf{b)}, the CISS molecules are aligned in-plane of the device, therefore we assume the ferromagnet also has in-plane magnetization, and we describe spin accumulation again as $\mu_s=\mu_{\rightarrow}-\mu_{\leftarrow}$. We simplify the discussion by assuming the spin injection is mainly contributed by the ferromagnet, and the spin detection is achieved through the spin-charge conversion mechanism of the CISS molecules. The spin accumulation underneath the detection contacts can be generated by two mechanisms: the spin diffusion in graphene (as in the previous case), and the spin-charge conversion between the injection node and the detection node. However, in the BCB model, the spin-charge conversion does not contribute to a spin accumulation underneath the detector contacts. Therefore, we only consider the spin diffusion mechanism. Similar to the previous case, we can derive the nonlocal signal
\begin{equation} \color{black}
R_{nl}=\pm k_{conv} P_{FM} R_{\lambda} e^{-\frac{d}{\lambda_s}},
\end{equation}
where $k_{conv}$ is the spin-charge conversion coefficient described before, but here it concerns the proximity-induced CISS effect in the graphene channel, rather than the CISS molecules themselves. Important to realize, due to the proximity effect, the diffusion length $\lambda_s$ and the spin resistance $R_{\lambda}$ of graphene may differ from the previous case.

\section{Generalized CISS Model}\label{app:general}

The BCB model is a simplified model where the spin-dependent characteristics of a CISS molecule are exclusively originated from an ideal CISS center. However, the assumption of having an ideal spin-flip core in a molecule may not be accurate. Therefore, we assume a general form of transmission and reflection matrices, 
\begin{equation}
\mathbb{T}^P_R=\begin{pmatrix}
a,c\\b,d
\end{pmatrix},~
\mathbb{R}^P_R=\begin{pmatrix}
A,C\\B,D
\end{pmatrix},
\end{equation}
where
\begin{subequations}\label{seqn:8}
	\begin{equation}\label{seqn:8a}
	0\leq a, b,c,d,A,B,C,D\leq 1,
	\end{equation}
	\begin{equation}\label{seqn:8b}
	a+b+A+B=c+d+C+D=1,
	\end{equation}
	\begin{equation}\label{seqn:8c}
	a+c > b+d.
	\end{equation}
\end{subequations}
Here the first restriction Eqn.~\ref{seqn:8a} comes from the fact that all matrix elements are probabilities. The second restriction Eqn.~\ref{seqn:8b} addresses that for each spin component, the sum of its probabilities of being transmitted and being reflected equals $1$. The third restriction Eqn.~\ref{seqn:8c} is the conceptual description of the CISS effect, which shows that a spin polarization arises after transmission through a CISS molecule. This restriction is similar to Eqn.~\ref{seqn:ciss}, but here we determine a sign for the polarization because we have assumed the chirality ($P$-type) of the molecule and the electron flow direction ($R$ for rightwards).

We still assume that the molecule has $C_2$ symmetry, so that the transmission and reflection matrices for reversed current can be written as 
\begin{subequations}
\begin{equation}
\mathbb{T}^P_L= \begin{pmatrix}
0&1\\
1&0
\end{pmatrix} \cdot \mathbb{T}^P_R \cdot \begin{pmatrix}
0&1\\
1&0
\end{pmatrix},
\end{equation}
\begin{equation}
\mathbb{R}^P_L= \begin{pmatrix}
0&1\\
1&0
\end{pmatrix} \cdot \mathbb{R}^P_R \cdot \begin{pmatrix}
0&1\\
1&0
\end{pmatrix}.
\end{equation}
\end{subequations}
The matrices for $AP$-type molecules can be derived according to Fig.~\ref{fig:matrices}. Before we proceed with the mathematical proof, we discuss the validity of these symmetry assumptions.

Since we only consider the electron transport in CISS molecules, the $C_2$ symmetry that we require only refers to the symmetry in electron transport. Furthermore, this symmetry is also a requirement of the linear regime. A highly asymmetric molecule can still be treated with our model, but only under the assumption that it is composed of a symmetric ($C_2$) part which contributes to CISS, and asymmetrically distributed normal barriers. In terms of the symmetry relations between the two chiralities, it is by definition that the two chiral enantiomers are exact mirror images of each other, and thereby select opposite spins with equal probability (when placed in the same circuit environment).

We apply the reciprocity theorem to a geometry similar to Fig.~\ref{fig:2Tafm} in the main text, but now the BCB molecule is replaced by a generalized CISS molecule. The general reciprocity theorem requires that the two-terminal resistance remains unchanged under magnetization reversal regardless of the polarization of the ferromagnet. Therefore, we assume the ferromagnet is $100\%$ polarized for convenience. Following the same steps as in the \textit{FM-BCB} geometry, the two terminal transmission probabilities considering two magnetization directions and two current directions become
\begin{subequations}
\color{black}
\begin{equation}
T_{21}(\Rightarrow)= a+b +(c+d)\frac{B}{1-D},
\end{equation}
\begin{equation}
T_{21}(\Leftarrow)= c+d+(a+b)\frac{C}{1-A},
\end{equation}
\begin{equation}
T_{12}(\Rightarrow)=b+d+(a+c)\frac{C}{1-D},
\end{equation}
\begin{equation}
T_{12}(\Leftarrow)=a+c+(b+d)\frac{B}{1-A},
\end{equation}
\end{subequations}
where the arrows indicate the magnetization direction of the ferromagnet, and the subscripts indicate the electron flow direction. The reciprocity theorem requires these four expressions to have the same value
\begin{equation}
T_{12}(\Rightarrow)=T_{12}(\Leftarrow)=T_{21}(\Rightarrow)=T_{21}(\Leftarrow)=T,
\end{equation}
which can be written as a vector equation
\begin{equation}
\begin{pmatrix}
1&1&\frac{B}{1-D}&\frac{B}{1-D}\\
\frac{C}{1-A}&\frac{C}{1-A}&1&1\\
\frac{C}{1-D}&1&\frac{C}{1-D}&1\\
1&\frac{B}{1-A}&1&\frac{B}{1-A}
\end{pmatrix}\cdot\begin{pmatrix}
a\\b\\c\\d
\end{pmatrix}=T\begin{pmatrix}
1\\1\\1\\1
\end{pmatrix}.
\end{equation}

In order for this equation to be self-consistent, relation $$A=D$$ is required. Consequently, we can derive
\begin{subequations}
\begin{equation}
b=c,
\end{equation}
\begin{equation}
a-d=C-B.
\end{equation}
\end{subequations}
Therefore, the generalized form of the transmission and reflection matrices of a CISS molecule is
\begin{equation}
\color{black}
\mathbb{T}^P_R=\begin{pmatrix}
a&b\\b&a(1-s)
\end{pmatrix},~
\mathbb{R}^P_R=\begin{pmatrix}
A&B+as\\B&A
\end{pmatrix},
\end{equation}
with $a+b+A+B=1$, and $s$ ($0\leq s \leq 1$) being a quantitative description of the \textit{strength} of the CISS effect in a molecule.

\begin{figure}[htb]
	\includegraphics[width=\linewidth]{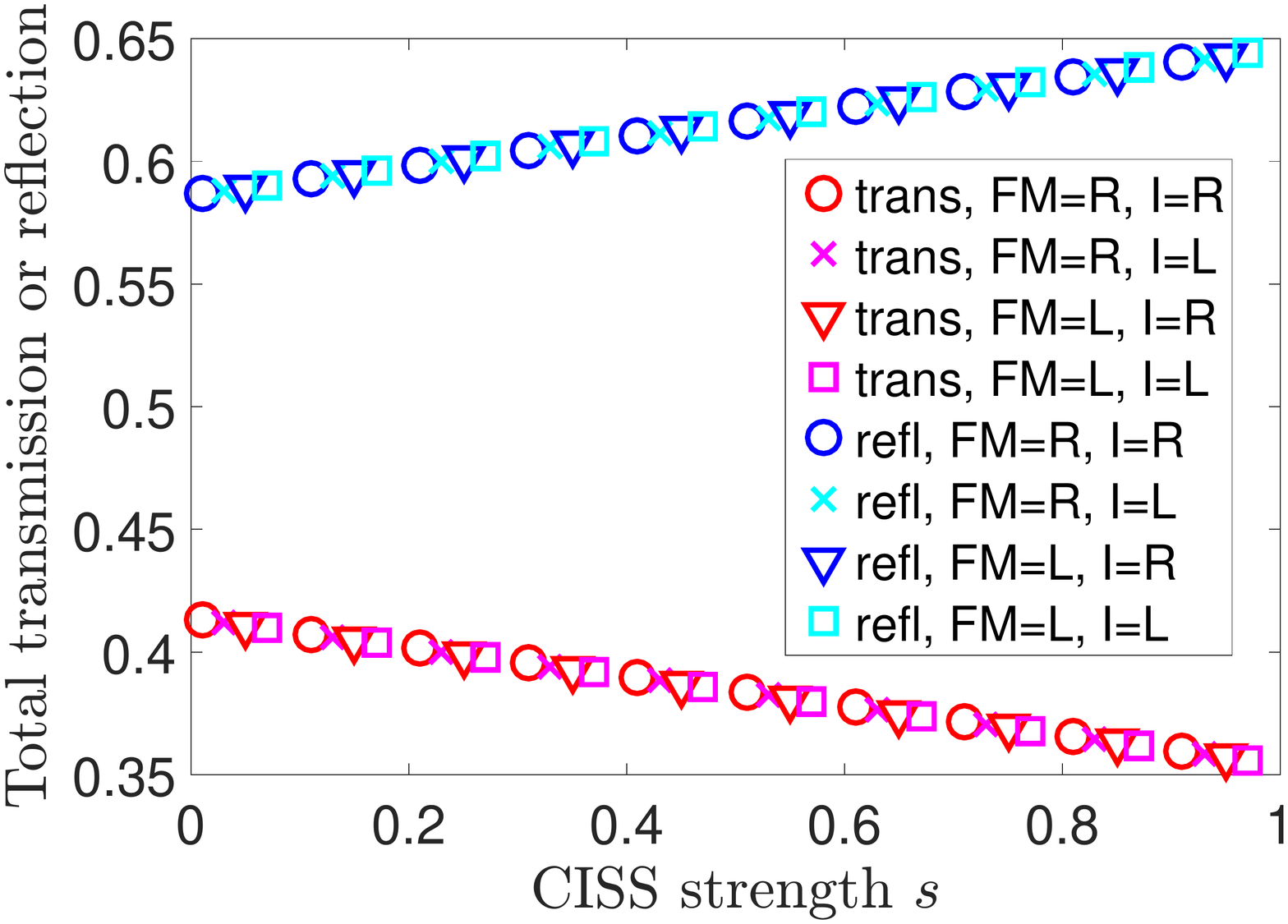}
	\caption{Normalized total transmission and reflection of an \textit{FM-CISS} segment as a function of $s$ (generalized CISS strength). Red and magenta labels (lower group) are for total transmissions accounting for both spin species for different magnetization orientations (FM=R or L) and different electron flow directions (I=R or L). Blue and cyan labels (upper group) are for total reflections. The polarization of the ferromagnet is chosen as $P_{FM}=0.7$ to be consistent with Fig.~\ref{sfig:cissF14}.}
	\label{sfig:cissF5}
\end{figure}

Next, we discuss the results given by this generalized model for the geometries discussed in the main text, and compare it with the BCB model. We arbitrarily choose the following transmission and reflection matrices for the generalized model
\begin{equation}
\color{black}
\mathbb{T}^P_R=\begin{pmatrix}
0.6&0.15\\
0.15&0.6(1-s)
\end{pmatrix},~  \mathbb{R}^P_R=\begin{pmatrix}
0.2&0.05+0.6s\\
0.05&0.2
\end{pmatrix},
\end{equation}
and use parameter $s$ as the variable to tune the \textit{CISS strength}.

For the \textit{FM-BCB} geometry (which now becomes \textit{FM-CISS} geometry), we plot the two-terminal transmission and reflection probabilities as a function of $s$ for four situations: two magnetization directions and two current directions. The results show that all four situations give identical results for all CISS strength $s$, as shown in Fig.~\ref{sfig:cissF5}. This is consistent with the BCB model, just as required by the reciprocity theorem.

\begin{figure}[htb]
	\includegraphics[width=\linewidth]{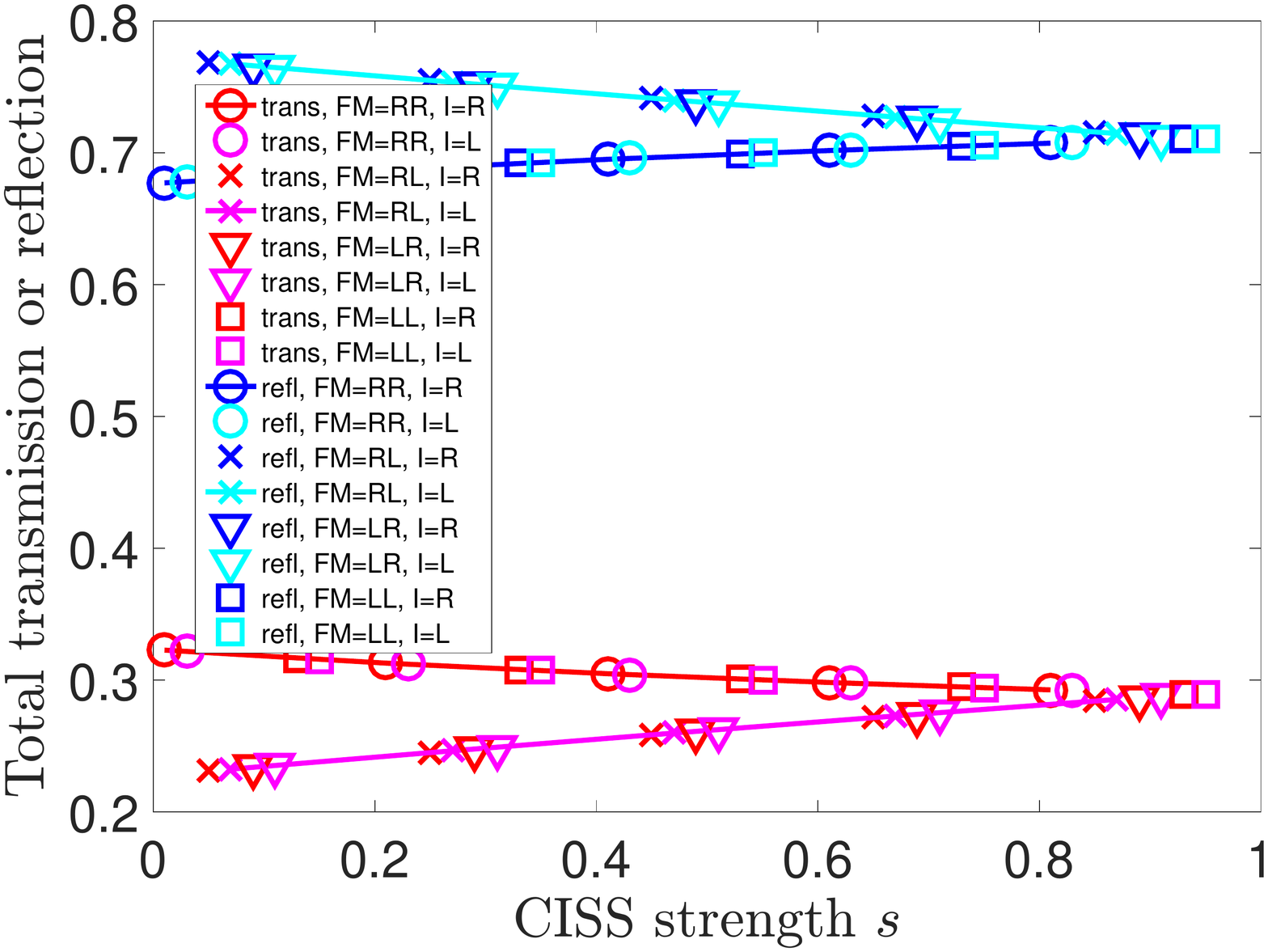}
	\caption{Normalized total transmission and reflection of a spin-valve segment as a function of $s$ (generalized CISS strength). Red and magenta labels (lower group) are for total transmissions accounting for both spin species for different ferromagnet configurations (two ferromagnets, FM=RR, RL, LR, or LL) and different current flow directions (I=R or L). Blue and cyan labels (upper group) are for total reflections. The polarization of the ferromagnet is chosen as $P_{FM}=0.7$ in order to amplify the deviation between parallel and anti-parallel ferromagnet configurations. This deviation is larger when the ferromagnet polarization is higher.}
	\label{sfig:cissF14}
\end{figure}

For the spin valve geometry, we plot the two-terminal transmission and reflection probabilities as a function of $s$ for eight situations: two ferromagnets each with two magnetization directions, and two current directions. The results are shown in Fig.~\ref{sfig:cissF14}. Unlike the BCB model, here we find different transmission and reflection probabilities for cases where the two magnetization directions are parallel \textit{vs.} anti-parallel. The transmission probability is in general higher when the two ferromagnets are magnetized parallel compared to anti-parallel. Notably, the difference between the two configurations decreases as the CISS strength $s$ increases. This is because with increasing $s$, electrons have a higher probability of encountering spin-flip reflections. This also explains why in a BCB model, where an ideal spin-flip process is present, switching the magnetization directions does not lead to any conductance variation. The direction of the current has no effect on the transmission and reflection probabilities, as is required by the linear regime.

\begin{figure}[htb]
	\includegraphics[width=\linewidth]{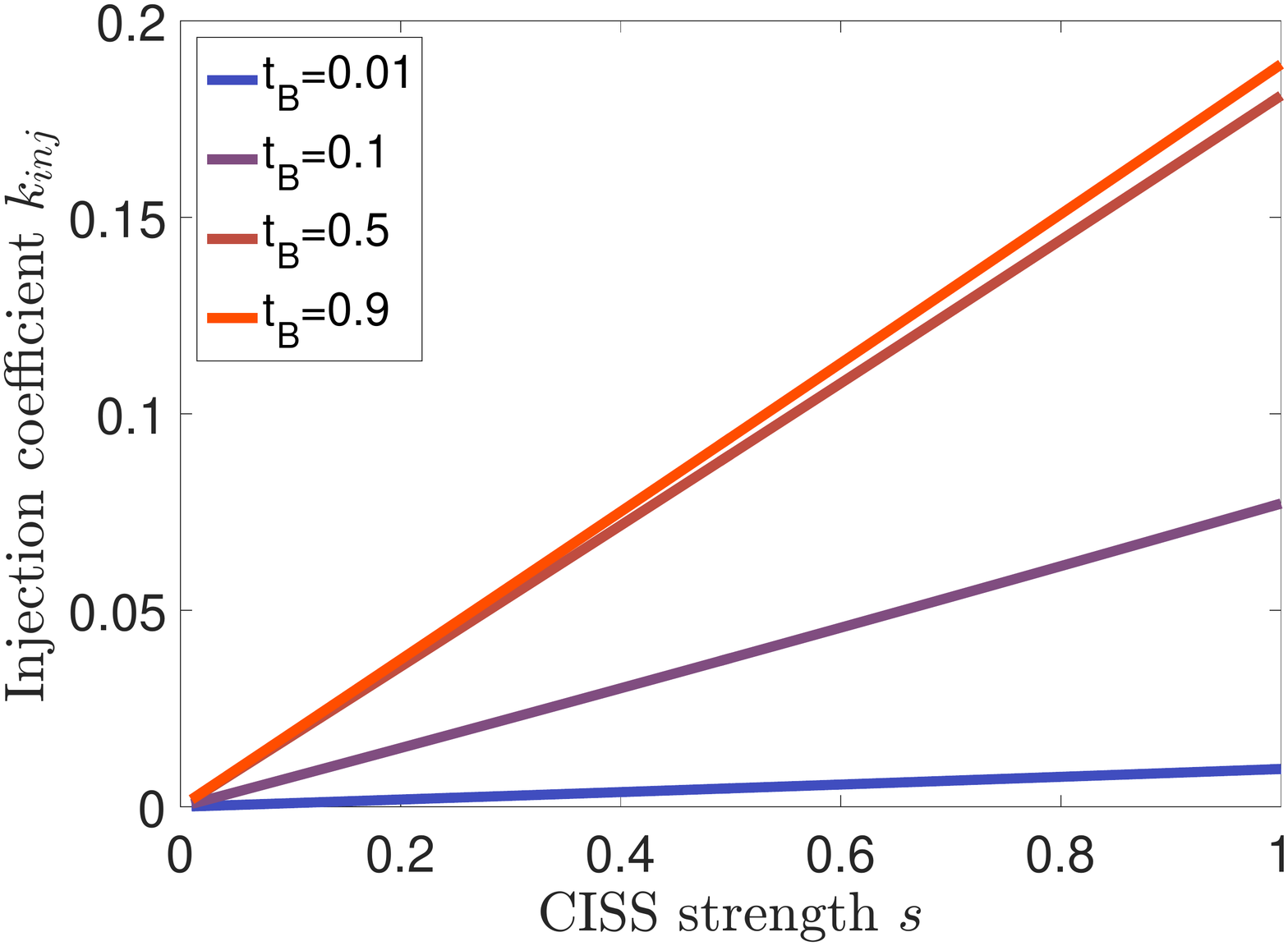}
	\caption{Injection coefficient $k_{inj}$ as a function of $s$ (generalized CISS strength) for various $t_B$ (contact barrier transmissiion), for the geometry described in Fig.~\ref{fig:4Tcross}\textbf{a)} in the main text.}
	\label{sfig:kinj}
\end{figure}

\begin{figure}[htb]
	\includegraphics[width=\linewidth]{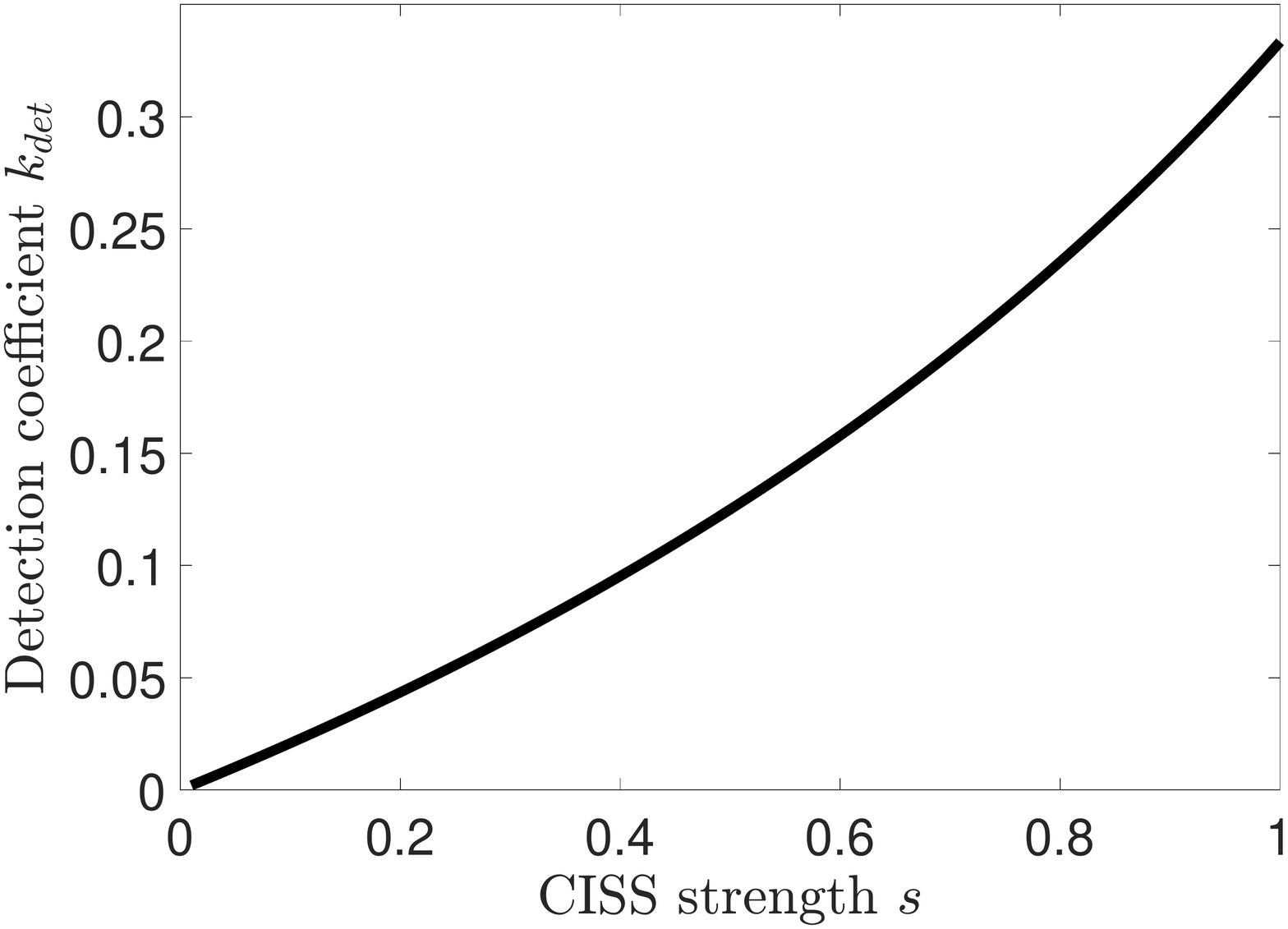}
	\caption{Detection coefficient $k_{det}$ as a function of $s$ (generalized CISS strength) for the geometry described in Fig.~\ref{fig:4Tcross}\textbf{a)} in the main text.}
	\label{sfig:kdet}
\end{figure}

\begin{figure}[htb]
	\includegraphics[width=\linewidth]{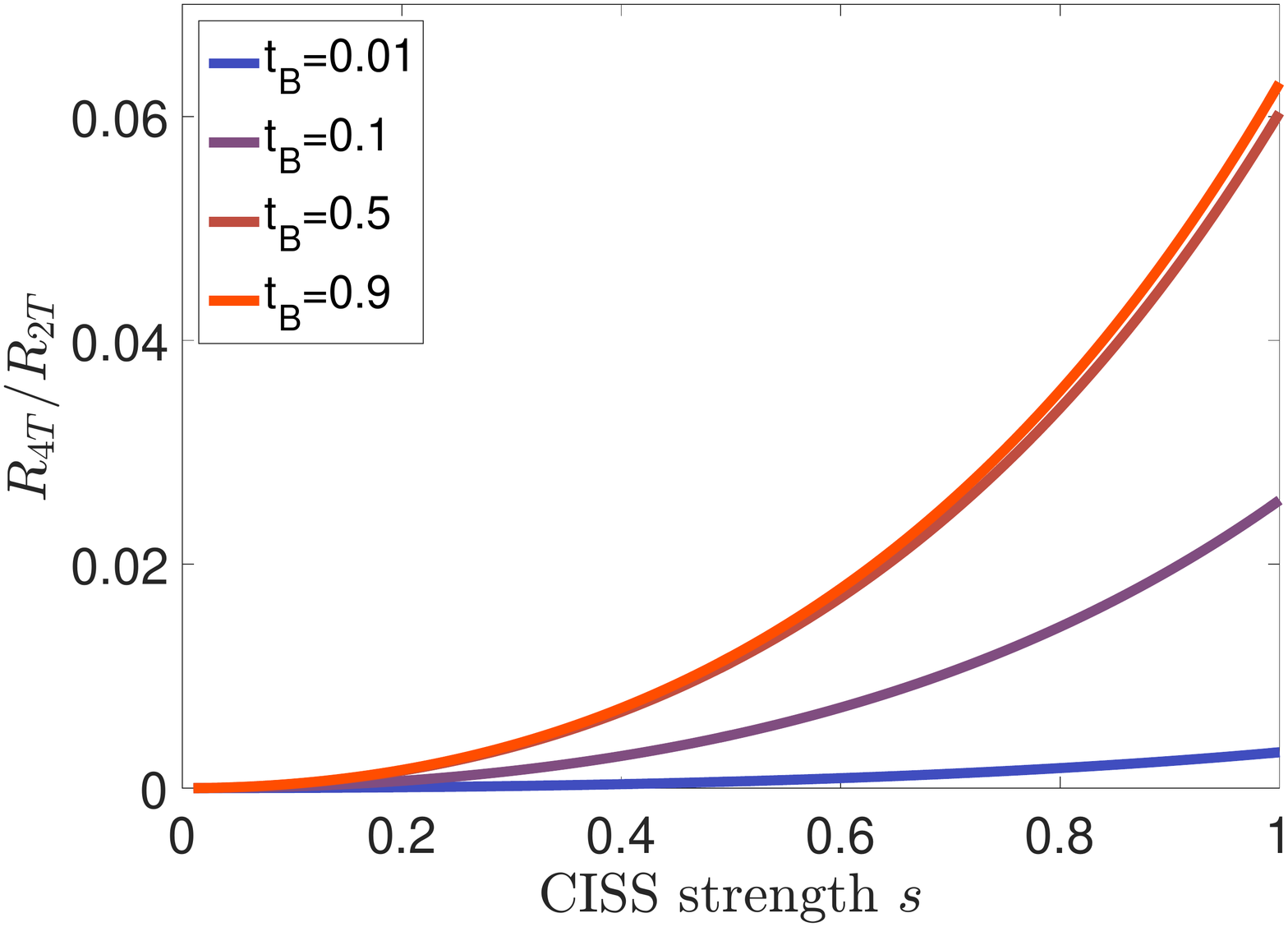}
	\caption{The ratio between four-terminal and two-terminal resistances as a function of $s$ (generalized CISS strength) for various $t_B$ (contact barrier transmission), for the geometry described by Fig.~\ref{fig:4Tcross}\textbf{a)} in the main text.}
	\label{sfig:kinjkdet}
\end{figure}

For four terminal geometries, we calculate here the spin injection and detection coefficients using the formulas derived in the main text, and we plot these coefficients as a function of the CISS strength $s$. Fig.~\ref{sfig:kinj} shows the injection coefficient as a function of $s$ for a few $t_B$ values. Fig.~\ref{sfig:kdet} shows the detection coefficient as a function of $s$, and Fig.~\ref{sfig:kinjkdet} shows the ratio between four-terminal and two-terminal resistances as a function of $s$. 

Last but not least, we discuss the spin-charge conversion property of the generalized CISS molecule. Substituting the corresponding transmission and reflection matrices into Eqn.~\ref{seqn:muns}, and solving the vector equation, we can obtain relations between the conversion coefficient $k_{conv}$ and the chemical potential vector $\boldsymbol{\mu_A}$. Unlike the BCB model, here it is not possible to drop out either $\mu_{nA}$ or $\mu_{sA}$ from the equation. The final result gives 
\begin{widetext}
\begin{equation}
k_{conv}=-\frac{\Delta \mu_n}{\Delta \mu_s}=\frac{ (A - B - 1) s \mu_{nA} +\frac{1}{2} (b + B) s \mu_{sA} }{2 (A+B-1) s \mu_{nA} +  (b + B) (-2 + 2 A + 2 B + s) \mu_{sA}},
\end{equation}
\end{widetext}
where $a,b,A,B,s$ are the parameters in the transmission and reflection matrices of the generalized CISS model. This equation shows that a non-zero charge voltage difference ($\Delta \mu_n$) can give rise to a spin accumulation difference ($\Delta \mu_s$) across a generalized CISS molecule, and \textit{vice versa} (spin-charge conversion). Interestingly, the conversion coefficient $k_{conv}$ depends not only on the transmission and reflection matrices of the molecule, but also on the spin accumulation in the nodes connected to the molecule. If $s=0$, i.e. the molecule does not exhibit any CISS effect, the spin-charge conversion property also diminishes ($k_{conv}=0$). 

With these discussions, we demonstrated that a generalized CISS molecule shows differences from the simplified BCB model. Nonetheless, these differences are rather quantitative than qualitative, and are not easily measurable in experiments. Furthermore, the generalized model also introduces extra degrees-of-freedom to calculations as it uses four variables to describe a CISS molecule, compared to one in the BCB model. Having taken the above into consideration, in the main text we only showed the results of the BCB model.

\cleardoublepage
\newpage

\begin{widetext}
\begin{center}
	\textbf{\large Supplementary information: additional figures}
\end{center}

\setcounter{equation}{0}
\setcounter{figure}{0}
\setcounter{table}{0}
\setcounter{page}{1}
\setcounter{section}{0}
\makeatletter
\renewcommand{\theequation}{S\arabic{equation}}
\renewcommand{\thefigure}{S\arabic{figure}}

Fig.~\ref{sfig:fmbcb1} to Fig.~\ref{sfig:fmbcb4} show the transmission and reflection probability of each spin component for each of the four situations (two magnetization directions and two current directions) of the \textit{FM-BCB} geometry. In these four situations, even though the total transmissions (or reflections) concerning both spin components are identical, the transmissions and reflections for each spin component, and hence the spin polarizations of the electrons entering the contact are different. This difference cannot be measured as a charge signal because the contacts cannot distinguish between spin components.

Fig.~\ref{sfig:sv1} to Fig.~\ref{sfig:sv8} show the same analysis for the eight situations (two ferromagnets each with two magnetization directions, and two current directions) of the spin valve geometry. Similarly, spin polarization can be generated at the contacts, and are different for all situations. However, due to the limitation of the two-terminal geometry, the total transmissions (or reflections) concerning both spin components remain the same for all eight situations. 

\begin{figure}[h!]
	\includegraphics[width=0.7\linewidth]{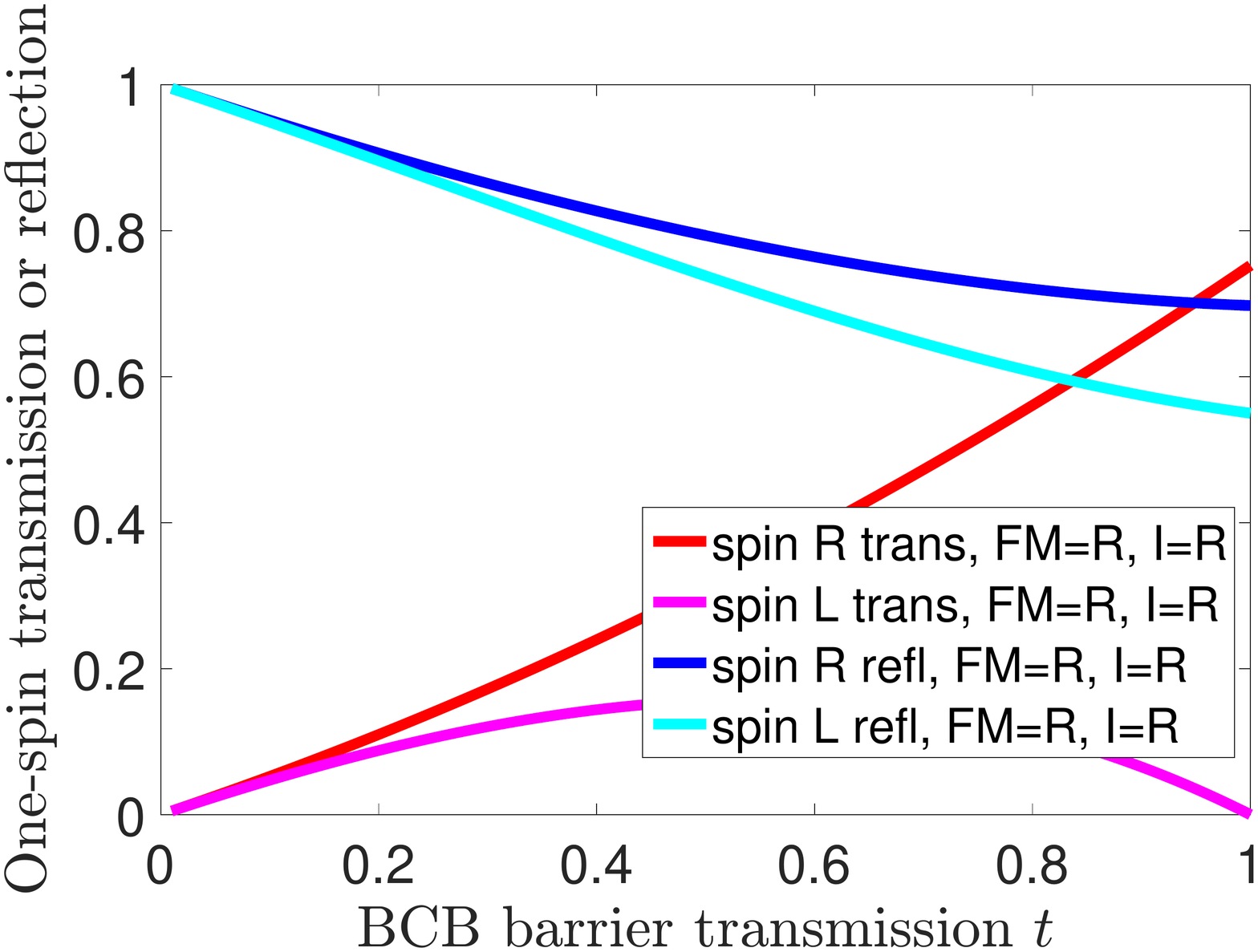}
	\caption{Normalized one-spin transmission or reflection for the FM-BCB case, where the magnetization of the ferromagnet is "R" (pointing to the right) and electron flow "R" (from left to right).}
	\label{sfig:fmbcb1}
\end{figure}
\begin{figure}[h!]
	\includegraphics[width=0.7\linewidth]{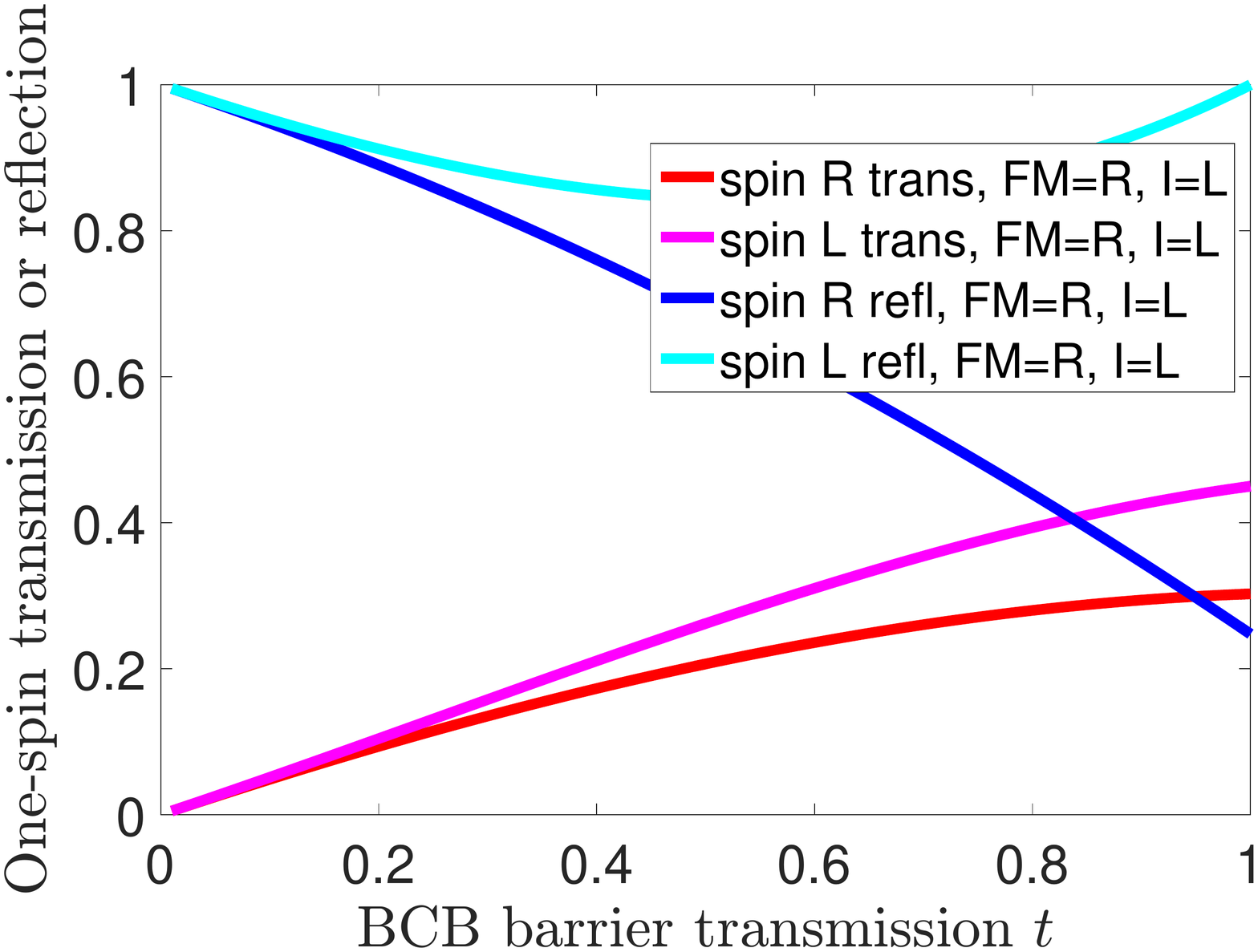}
	\caption{Normalized one-spin transmission or reflection for the FM-BCB case, where the magnetization of the ferromagnet is "R" (pointing to the right) and electron flow "L" (from right to left).}
\end{figure}
\begin{figure}[h!]
	\includegraphics[width=0.7\linewidth]{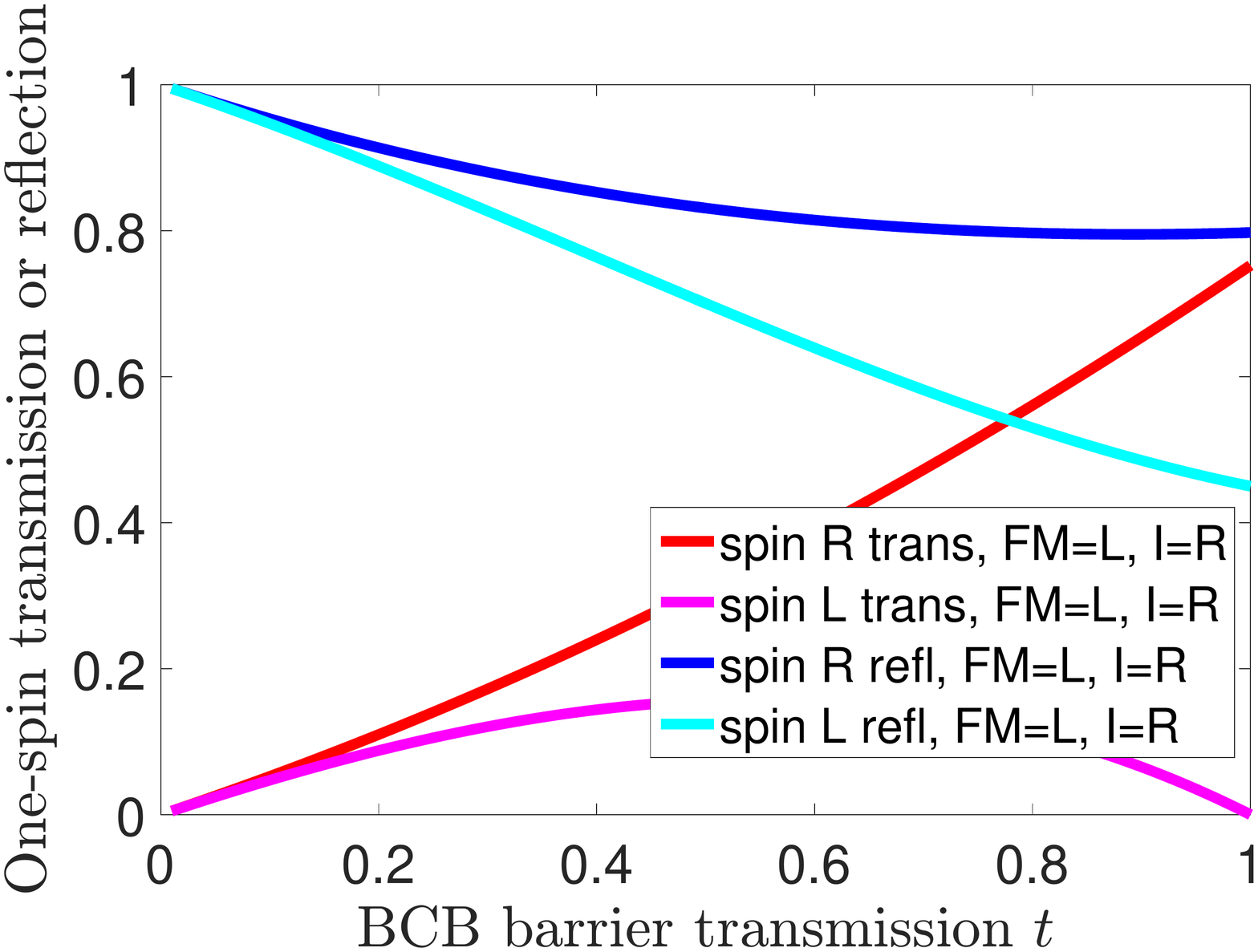}
	\caption{Normalized one-spin transmission or reflection for the FM-BCB case, where the magnetization of the ferromagnet is "L" (pointing to the left) and electron flow "R" (from left to right).}
\end{figure}
\begin{figure}[h!]
	\includegraphics[width=0.7\linewidth]{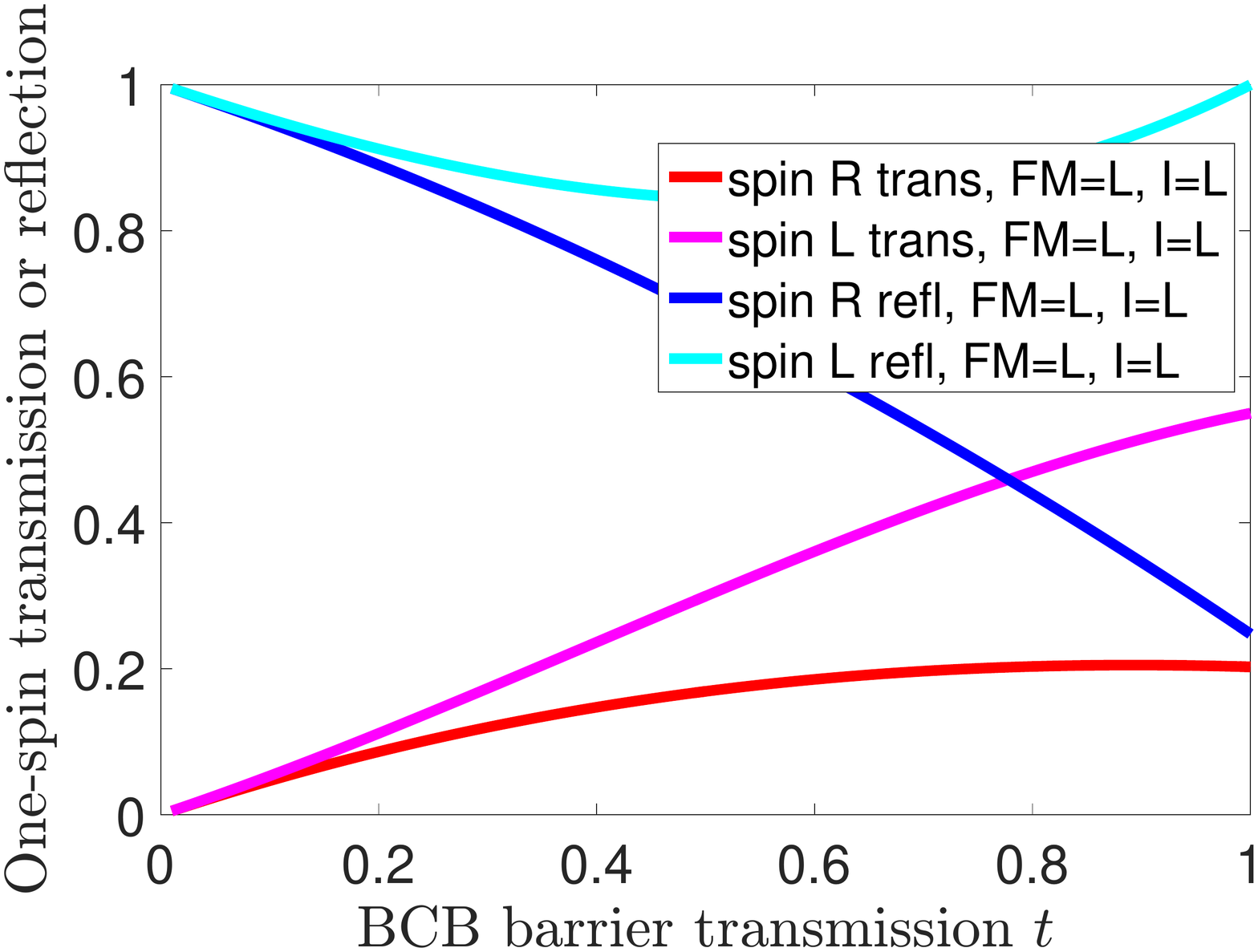}
	\caption{Normalized one-spin transmission or reflection for the FM-BCB case, where the magnetization of the ferromagnet is "L" (pointing to the left) and electron flow "L" (from right to left).}
	\label{sfig:fmbcb4}
\end{figure}

\begin{figure}[h!]
	\includegraphics[width=0.7\linewidth]{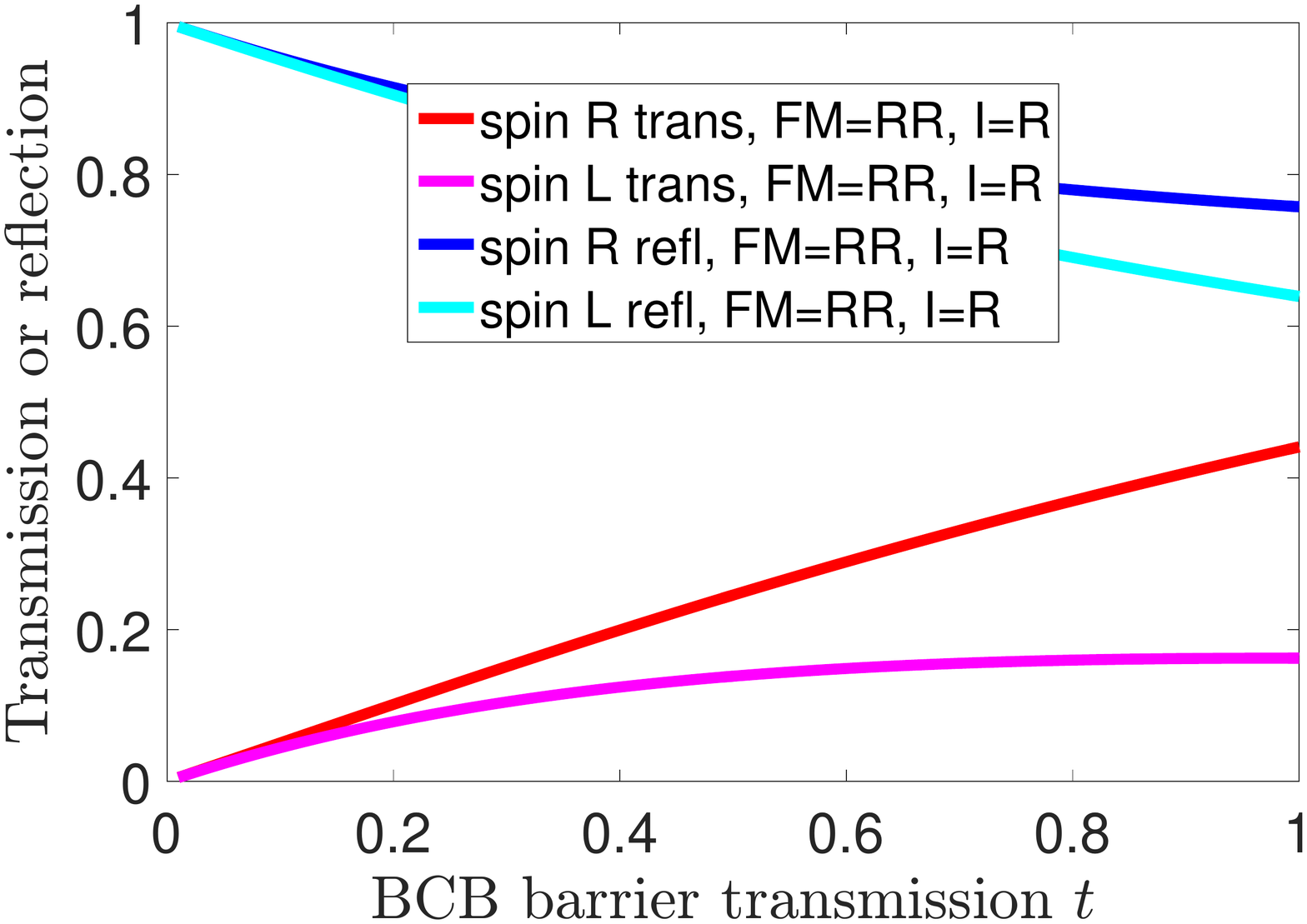}
	\caption{Normalized one-spin transmission or reflection for the spin valve case, where the magnetization of the ferromagnet is "RR" (both pointing to the right) and electron flow "R" (from left to right).}
	\label{sfig:sv1}
\end{figure}
\begin{figure}[h!]
	\includegraphics[width=0.7\linewidth]{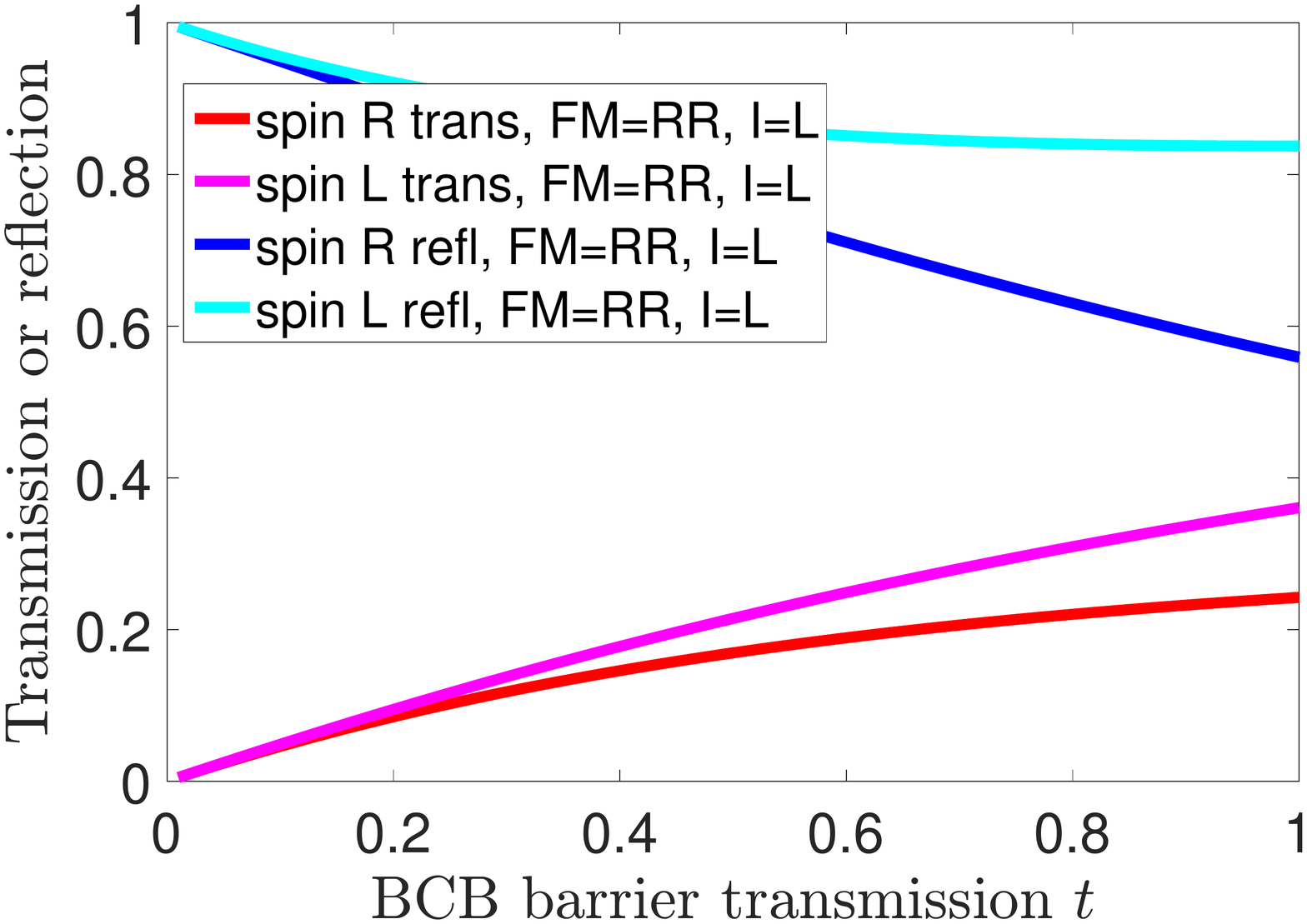}
	\caption{Normalized one-spin transmission or reflection for the spin valve case, where the magnetization of the ferromagnet is "RR" (both pointing to the right) and electron flow "L" (from right to left).}
\end{figure}
\begin{figure}[h!]
	\includegraphics[width=0.7\linewidth]{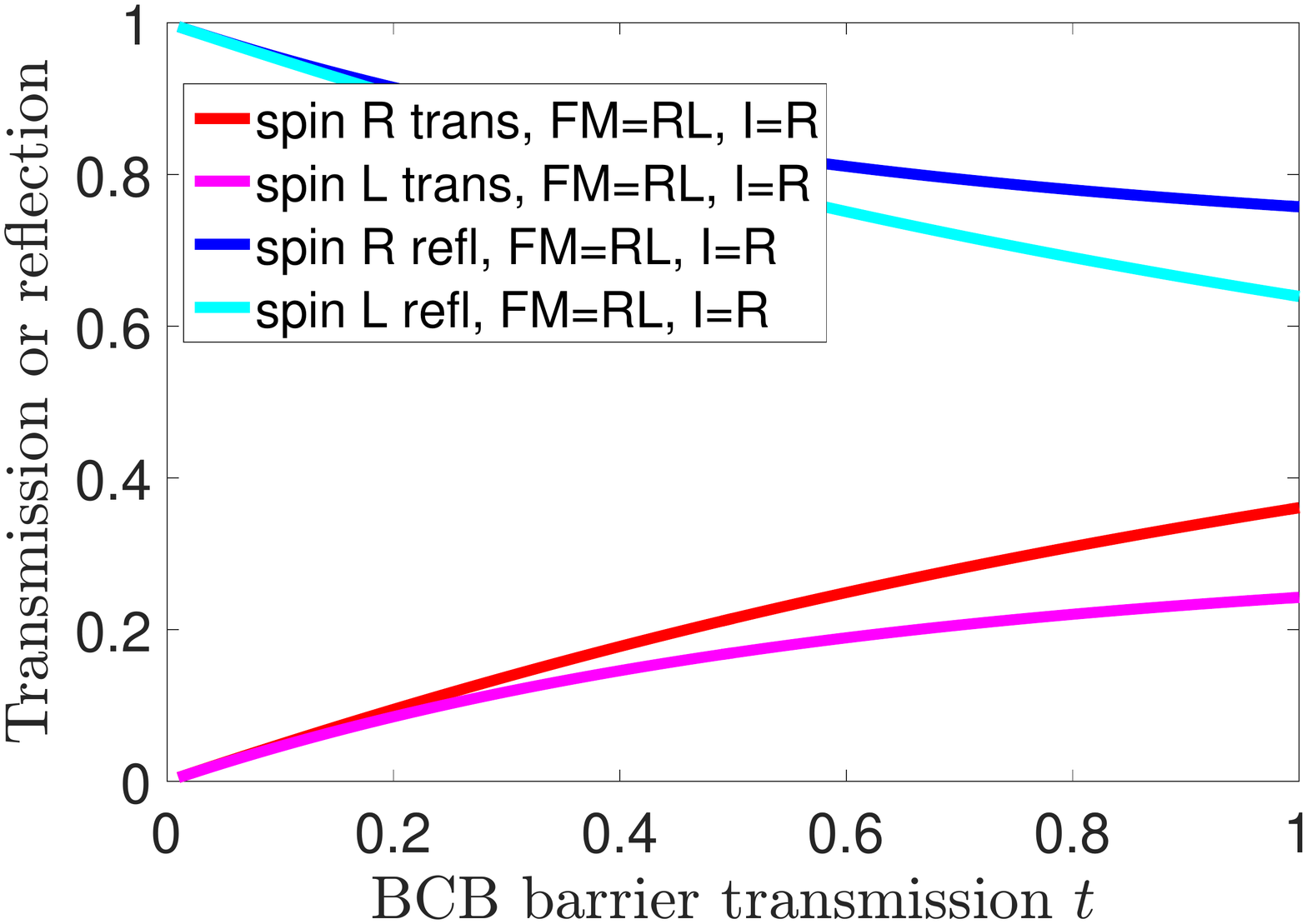}
	\caption{Normalized one-spin transmission or reflection for the spin valve case, where the magnetization of the ferromagnet is "RL" (the first one pointing to the right, and the second one to the left) and electron flow "R" (from left to right).}
\end{figure}
\begin{figure}[h!]
	\includegraphics[width=0.7\linewidth]{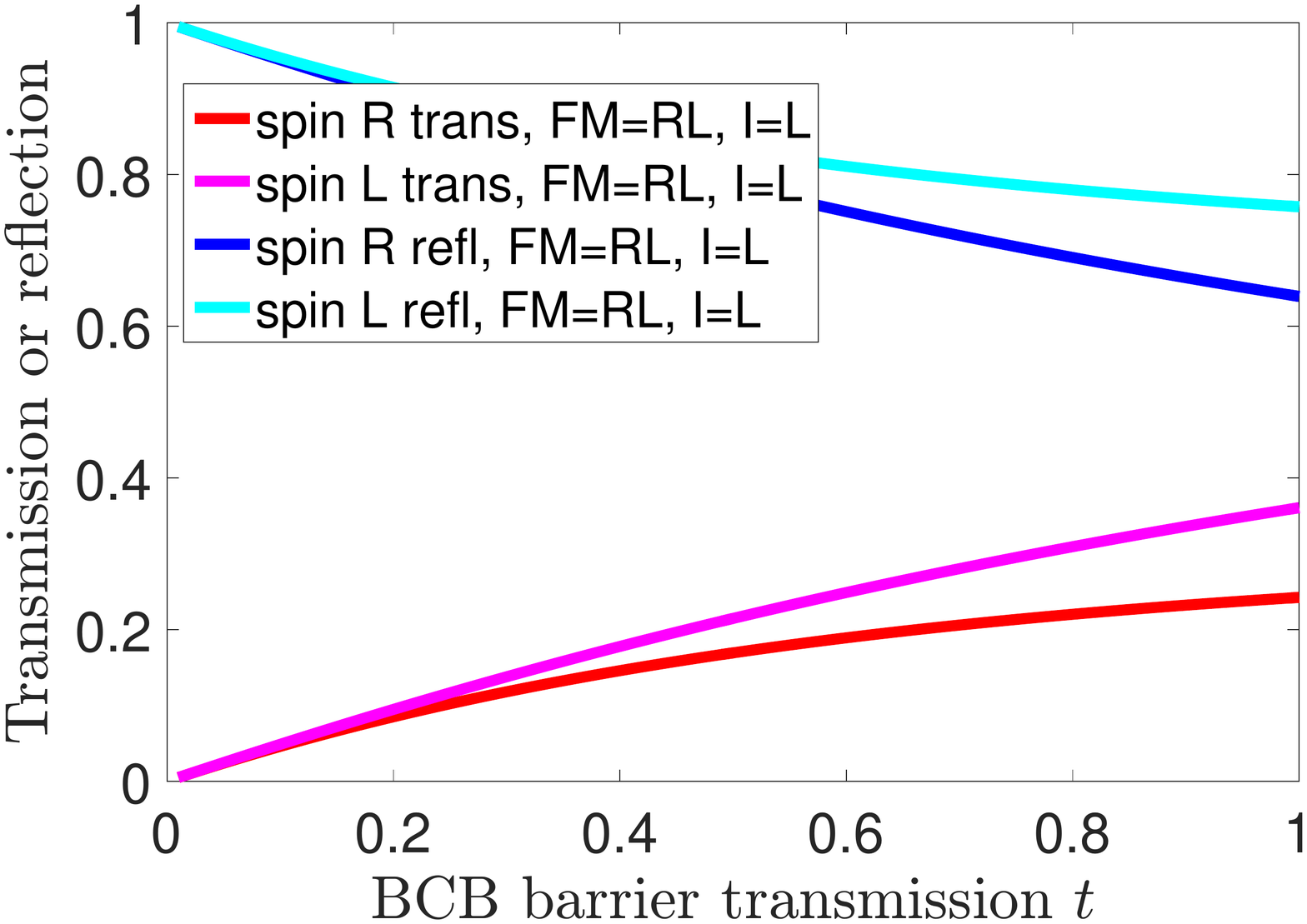}
	\caption{Normalized one-spin transmission or reflection for the spin valve case, where the magnetization of the ferromagnet is "RL" (the first one pointing to the right, and the second one to the left) and electron flow "L" (from right to left).}
\end{figure}

\begin{figure}[h!]
	\includegraphics[width=0.7\linewidth]{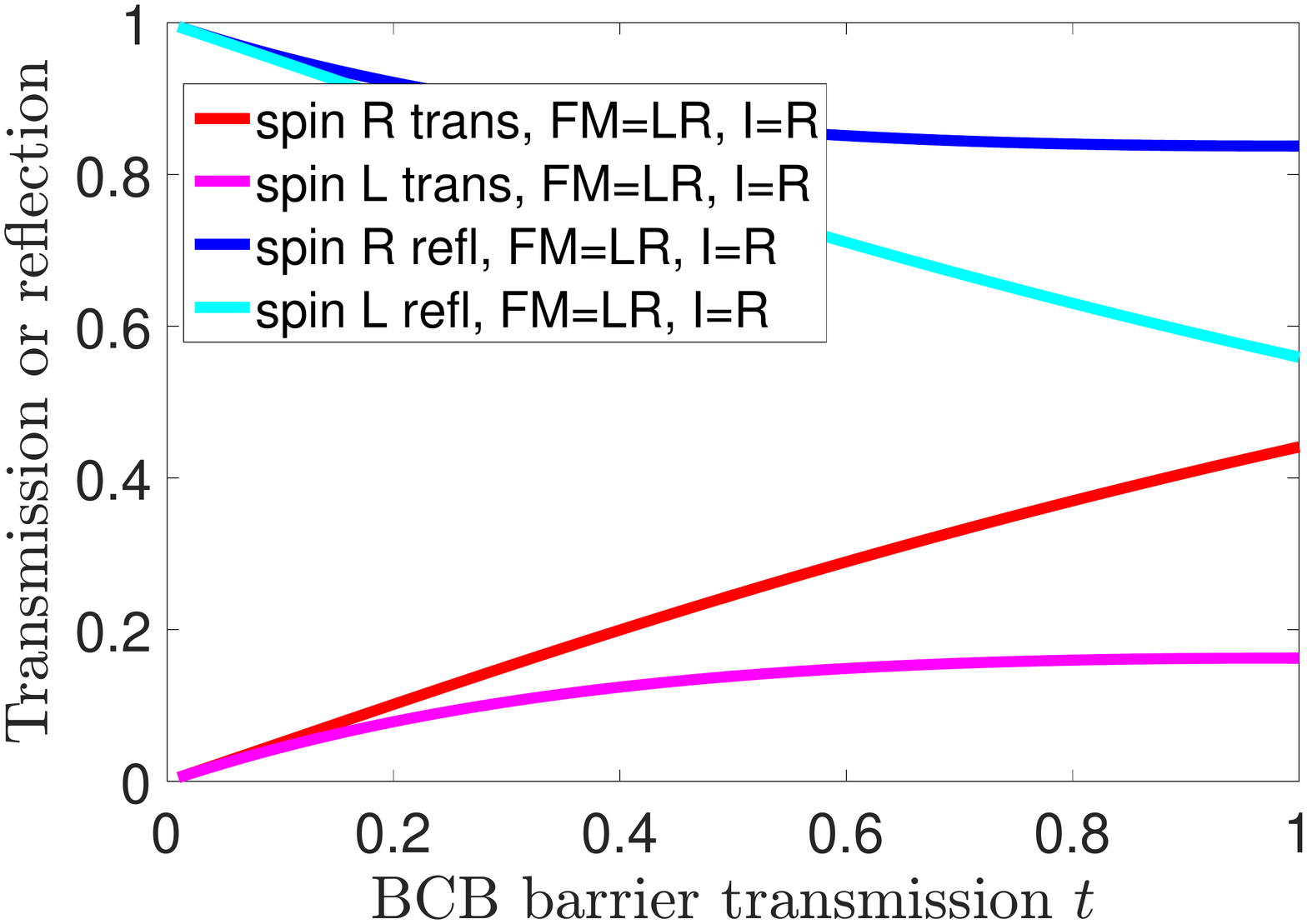}
	\caption{Normalized one-spin transmission or reflection for the spin valve case, where the magnetization of the ferromagnet is "LR" (the first one pointing to the left, and the second one to the right) and electron flow "R" (from left to right).}
\end{figure}
\begin{figure}[h!]
	\includegraphics[width=0.7\linewidth]{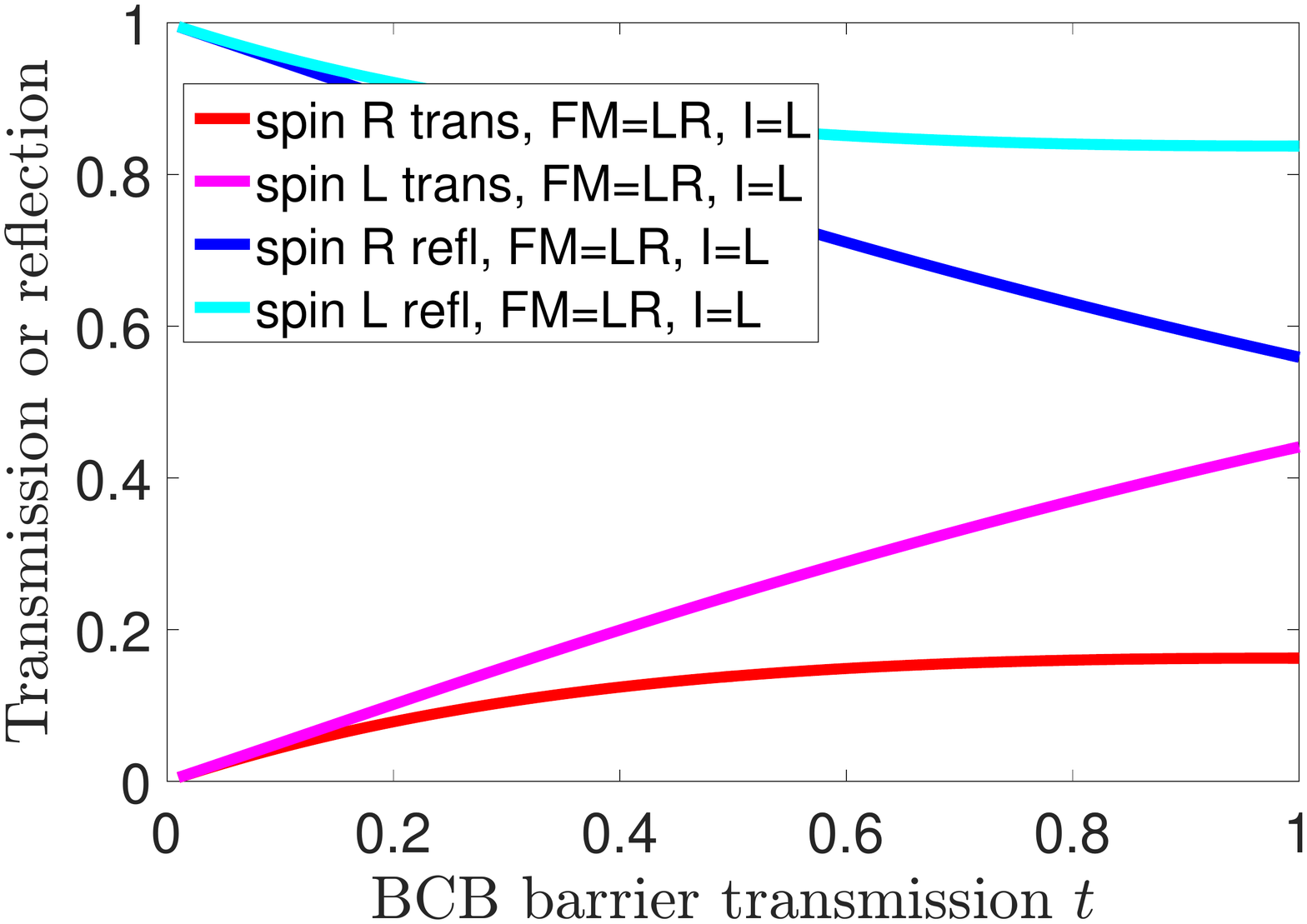}
	\caption{Normalized one-spin transmission or reflection for the spin valve case, where the magnetization of the ferromagnet is "LR" (the first one pointing to the left, and the second one to the right) and electron flow "L" (from right to left).}
\end{figure}
\begin{figure}[h!]
	\includegraphics[width=0.7\linewidth]{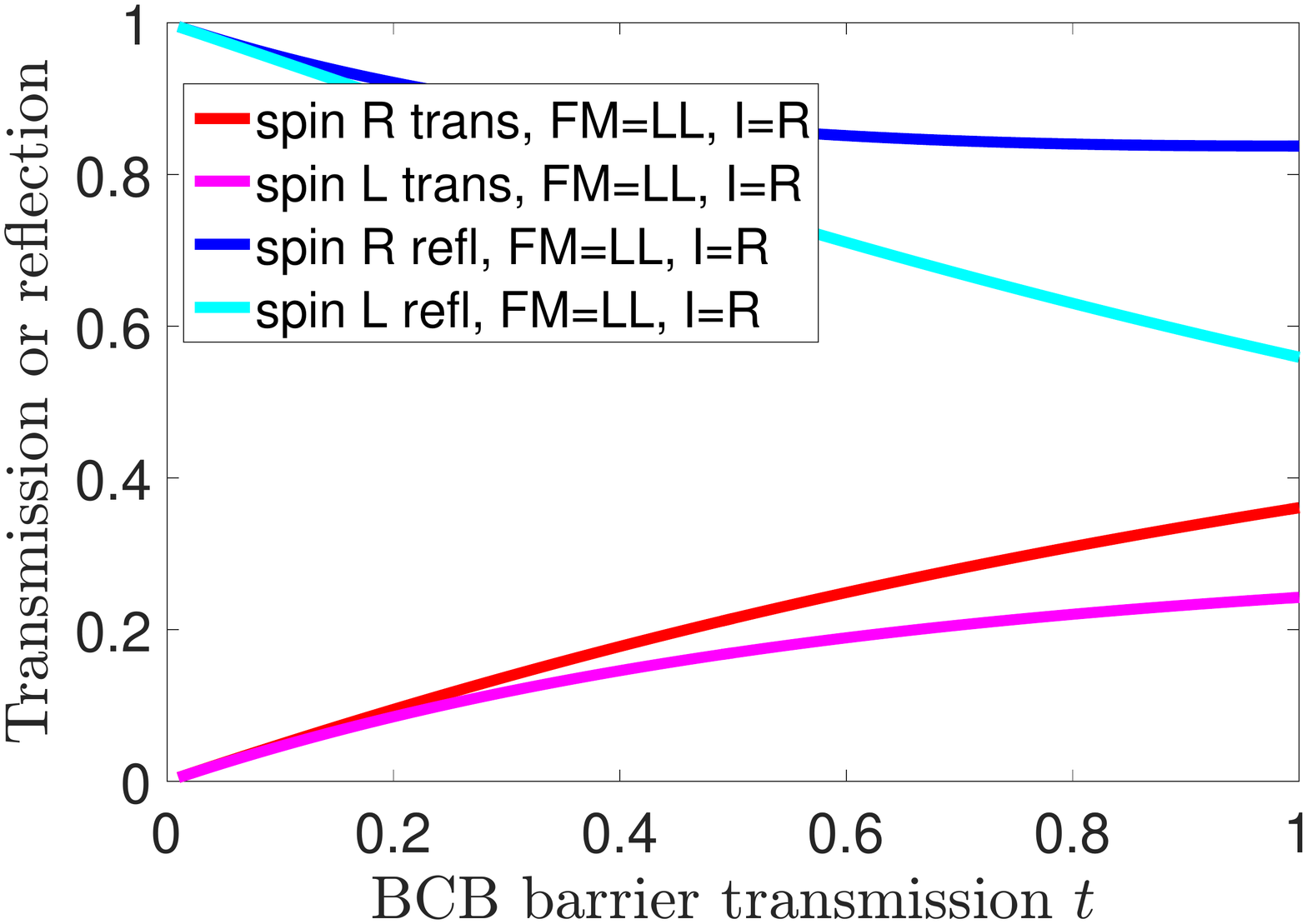}
	\caption{Normalized one-spin transmission or reflection for the spin valve case, where the magnetization of the ferromagnet is "LL" (both pointing to the left) and electron flow "R" (from left to right).}
\end{figure}
\begin{figure}[h!]
	\includegraphics[width=0.7\linewidth]{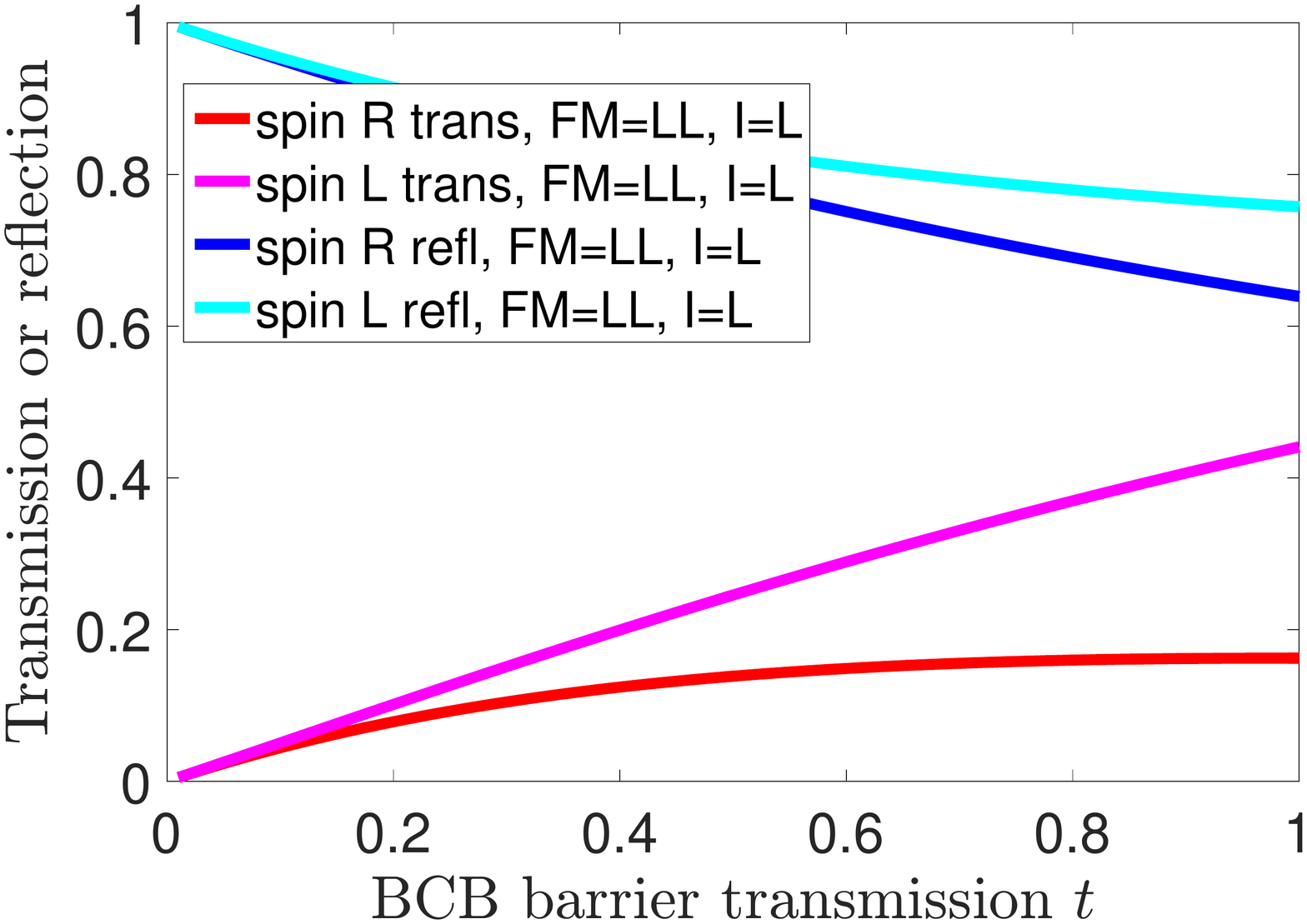}
	\caption{Normalized one-spin transmission or reflection for the spin valve case, where the magnetization of the ferromagnet is "LL" (both pointing to the left) and electron flow "L" (from right to left).}
	\label{sfig:sv8}
\end{figure}
\end{widetext}

\end{document}